\documentclass[3p,preprint]{elsarticle}
\usepackage{amssymb}
\usepackage{amsthm}
\usepackage{amsmath}
\usepackage[dvips]{epsfig}
\usepackage{graphicx}
\usepackage{color}
\usepackage{url}
\usepackage{bm,upgreek}
\usepackage{caption}
\usepackage{caption,setspace}
\usepackage{subcaption}
\usepackage{epstopdf}
\usepackage{multirow}
\usepackage{hhline}
\usepackage{slashbox}
\usepackage{cuted}

\usepackage[normalem]{ulem}

\newcommand{\bh}{{\bf h}}
\newcommand{\bu}{{\bf u}}
\newcommand{\bA}{{\bf A}}
\newcommand{\bB}{{\bf B}}
\newcommand{\PT}{{\cal PT}}

\newcommand{\tx}{\tilde{x}}

\begin{document}

%%%% Article title to be placed here
\title{Global search for localised modes in  scalar and vector nonlinear Schr\"odinger-type equations}

\author[miet,ufa]{G.L. Alfimov}
\ead{galfimov@yahoo.com}

\author[du,uc,ba]{I.V.  Barashenkov}

\author[miet]{A.P.  Fedotov}
%\ead{mathpi314159@yandex.ru}

\author[miet]{V.V. Smirnov}
%\ead{mathpi314159@yandex.ru}

\author[itmo]{D. A. Zezyulin}
%\ead{dzezyulin@corp.ifmo.ru}

\address[miet]{National Research University of Electronic Technology MIET, Zelenograd, Moscow 124498, Russia}

\address[ufa]{Institute of Mathematics with Computer Center, Ufa Scientific Center, Russian Academy of Sciences, Chernyshevskii~str. 112, Ufa 450008, Russia}

\address[du]{Joint Institute for Nuclear Research, Dubna 141980, Russia}

\address[uc]{
 {Department of Mathematics, University of Cape Town, Rondebosch 7701,
  South Africa}}

\address[ba]{Department of Physics, University of Bath, Claverton Down,
Bath BA2 7AY, UK}

\address[itmo]{ITMO University, St. Petersburg 197101, Russia}

%%%%%%%%% Insert author address here
%\address{$^{1}$ Moscow Institute of Electronic Engineering,
%    Zelenograd, Moscow, 124498, Russia\\
%$^{2}$Centro de F\'isica Te\'orica e Computacional and Departamento de
%F\'isica, Faculdade de Ci\^encias, Universidade de Lisboa, Avenida
%Professor Gama Pinto 2, Lisboa 1649-003, Portugal}

%%%% Subject entries to be placed here %%%%
%\subject{waves motion}

%%%% Keyword entries to be placed here %%%%
%\keywords{nonlinear mode, soliton, collapsing solution, nonlinear Schr\"odinger equation, Gross-Pitaevskii equation, Lugiato-Lefever equation}

%%%% Insert corresponding author and its email address}
%\corres{Georgy L. Alfimov\\
%\email{galfimov@yahoo.com}}

%%%% Abstract text to be placed here %%%%%%%%%%%%

%%%%%%%%%%%%%%%%%%%%%%%%%%%

%%%%%%%%%% Insert the texts which can accomdate on firstpage in the tag "fmtext" %%%%%

%\begin{fmtext}
%\end{fmtext}
%%%%%%%%%%%%%%% End of first page %%%%%%%%%%%%%%%%%%%%%
\begin{abstract}
We present a new approach for search of {coexisting classes of}
 localised modes admitted by the repulsive (defocusing)
 scalar or vector  nonlinear Schr\"odinger-type equations.
The approach is based on  the observation that
generic solutions of the corresponding stationary system
have  singularities at finite points on the real axis. We start with establishing conditions on the initial data of the associated Cauchy problem that guarantee the  formation of a singularity.
Making use of these sufficient conditions, we {identify} the bounded, nonsingular, solutions ---
and then classify them according to their asymptotic behaviour.
To determine the bounded solutions,  a properly chosen space of initial data is scanned numerically.
Due to asymptotic or symmetry considerations, we can limit ourselves to  a one- or two-dimensional space.
{For each set of initial conditions we compute the distances $X^{\pm}$ to the nearest forward and backward singularities;
 large $X^+$ or $X^-$ indicate the proximity to a bounded solution.}
We illustrate  our method with the Gross-Pitaevskii equation with a $\PT$-symmetric complex potential,
a system of coupled Gross-Pitaevskii equations with real potentials, and the  Lugiato-Lefever equation with normal dispersion.
\end{abstract}


\maketitle

\section{Introduction}
\label{sec:intro}

{The aim of this paper is to formulate a new approach to the determination of  localised solutions of the nonlinear Schr\"odinger-type systems.
The method is particularly efficient in  situations where  localised solutions are not unique; it aims at finding {\it all\/} solutions coexisting for
a given set of parameter values. The new approach
 applies to a broad variety of scalar
and vector equations, autonomous or not.
 We start with providing several examples of physical significance.}

1. The {\it Gross-Pitaevskii}  equation,
\begin{gather}\label{GP}
i\psi_t+\psi_{xx}-V(x)\psi-|\psi|^2\psi=0,
\end{gather}
describes the ground state of the quantum system of identical bosons in the mean field approximation \cite{PS03}. Equation  (\ref{GP}) is written in its dimensionless form, and corresponds to the elongated (``cigar-shaped'') condensate with repulsive interparticle interactions. In (\ref{GP}), $V(x)=V_1(x)+iV_2(x)$ is a complex-valued function whose real part $V_1(x)$
represents the confining potential. The imaginary part $V_2(x)$ accounts for the  injection and elimination of atoms in the condensate in the region where $V_2(x)$ is positive and negative, respectively.

Assuming a steady state solution of the form   $\psi(x,t)=e^{-i\mu t}u(x)$,   equation (\ref{GP}) reduces to  a nonlinear ordinary differential equation
\begin{gather}
u_{xx}+(\mu-V(x))u-|u|^2u=0.\label{GP_S}
\end{gather}
It is well-known that even in the case of  {a}  real potential $V(x)=V_1(x)$ (parabolic or periodic) and real amplitude $u(x)$,  equation (\ref{GP_S}) can describe numerous nonlinear structures.  {These include} bright and dark solitons \cite{EdwBur1995,Rupr1995,DalfString1996,Kivshar2001,MalomLasPhys2002,AgPres2002, Kiv2003,Dark,AKS,PelKiv,AlfZez07},  nonlinear periodic structures
\cite{Kiv2003,BlochW}, domain walls
\cite{DomWalls05}, gap waves \cite{Kiv2006} and even more complex multi-soliton structures that can be classified by means of some coding procedure \cite{AlfKizZez17}.

{The}  situation becomes  {even more involved when the potential $V(x)$ is complex.}
One of the notable realisations
{here is concerned with}   condensates in parity-time ($\PT$)-symmetric potentials   \cite{PTGPE01,PTGPE02,PTGPE03,KYZ16}. In this case $V_1(x)$ is an even function, i.e.,  $V_1(-x)=V_1(x)$ and    $V_2(x)$ is odd,  $V_2(-x)=-V_2(x)$. Another area where equation  (\ref{GP}) with a $\PT$-symmetric potential finds  applications, is nonlinear optics. In optics, $V(x)$ describes the complex-valued refractive index of a defocusing waveguide, where domains with positive and negative imaginary part of $V(x)$ correspond to  gain and loss of energy, respectively \cite{PTOptics1}.
A wide variety of optical potentials has been considered, see e.g. \cite{PTOptics2,PTOptics3,PTOptics4,PTOptics5,PTOptics6,ZK12} for particular examples and \cite{KYZ16,PTOptics7,PTGPE04} for  recent reviews.

Decomposing $u(x)$ into its real and imaginary parts,  $u(x)=u_1(x)+iu_2(x)$, we write (\ref{GP_S}) as a system
\begin{align}
&u_{1,xx}+(\mu-V_1(x))u_1+V_2(x)u_2-\left(u_1^2+u_2^2\right)u_1=0,\label{GP01}\\
&u_{2,xx}-V_2(x)u_1+(\mu-V_1(x))u_2-\left( u_1^2+ u_2^2\right)u_2=0.\label{GP02}
\end{align}
Complete description of all coexisting localised modes described by (\ref{GP01})-(\ref{GP02}) is a complex problem that generically remains unsolved.\medskip

2.  The dynamics of {\it a mixture} of two Bose-Einstein condensates  (without the injection or elimination of atoms)
is described by the coupled Gross-Pitaevskii equations
\begin{align}
&i\psi_{1,t}+\psi_{1,xx}-V(x)\psi_1-(\beta_{11}|\psi_1|^2+\beta_{12}|\psi_2|^2)\psi_1=0,\label{2_GP01}\\
&i\psi_{2,t}+\psi_{2,xx}-V(x)\psi_2-(\beta_{12}|\psi_1|^2+\beta_{22}|\psi_2|^2)\psi_2=0.  \label{2_GP02}
\end{align}
(See, e.g. the recent review \cite{KF16}). Here $\psi_{1,2}(x,t)$ are the macroscopic wavefunctions of the condensates, $\beta_{mn}$ characterize the inter-atomic collisions, and
the real function $V(x)$ describes the trap potential.
The generalisation to the case of three or more condensates is straightforward.

It is worth noting that equations (\ref{2_GP01})-(\ref{2_GP02}) emerge  in the optical context as well.
In optics, this system describes the propagation of two light beams in the presence of cross-phase modulation  \cite{Agrawal}.

The list of known localised solutions  of this system comprises the nodeless vector solitons  (aka ground states or bright-bright solitons)
  \cite{HoSh96,PB98,NavKevCarretero09};
their  dark-bright  \cite{C88,AKKS89,BA01,KevrSurv16,DB_17},
dark-dark   \cite{BrazhKon05,HCHE11,YCHK12},  and dark-antidark  \cite{KNFMC04} counterparts,
as well as localised solutions with more complex structure \cite{NavKevCarretero09}.

One of the applications of the system (\ref{2_GP01})-(\ref{2_GP02}) is to the description of  the
{\it miscibility-immiscibility transition} (MIT) in a binary Bose--Einstein  condensate.
As the inter-species coefficient
$\beta_{12}$ is increased through a critical value, the mixed components  separate.
A simple qualitative estimate for the threshold  is  $\beta_{12}      =\sqrt{     \beta_{11}\beta_{22}}$ \cite{ACh98,TGRMB_00} ---
although this expression  is known not to be exact \cite{Condition_Chin12}.

The substitution
$\psi_{1,2}(x,t)=e^{-i\mu_{1,2}t}u_{1,2}(x)$ with real $u_{1,2}(x)$, takes  (\ref{2_GP01})-(\ref{2_GP02}) to  the
following system:
\begin{align}
u_{1,xx} &+ (\mu_1-V(x))u_1 -
(\beta_{11}  u_1 ^{2}+\beta_{12}  u_2 ^{2}) u_1=0,
\label{Stat_Sch_Gen01}\\[2mm]
u_{2,xx} &+ (\mu_2-V(x))u_2 -
(\beta_{12}  u_1^{2}+\beta_{22}  u_2^{2}) u_2=0.
\label{Stat_Sch_Gen02}
\end{align}
%From mathematical viewpoint
The MIT transition corresponds to the symmetry-breaking bifurcation of its  ground state solution. In order to  seek for new localised solutions and  study bifurcations of the known ones,
it is useful to have
a global picture of the phase space of the system   \eqref{Stat_Sch_Gen01}-\eqref{Stat_Sch_Gen02} at a given set of parameter values. This global view should include all nonlinear modes determined so far.

\medskip

3. The externally driven nonlinear Schr\"odinger  equation (also known  as the {\it  Lugiato-Lefever}  equation),
\begin{gather}\label{LL2}
i \psi_t+ \frac12\psi_{xx}-\psi \pm|\psi|^2\psi=-i\gamma \psi + h,
\end{gather}
% has been studied since the mid 80s  %starting with pioneering work of L.A.Lugiato and R.Lefever
%\cite{LL87}. Equation~(\ref{LL2}). It
arises as an amplitude equation for an optical pulse in the pumped ring resonator \cite{LL87}.  Here $\psi(x,t)$ is a complex amplitude of the field in the resonator, while $\gamma$ and $h$ stand for the damping coefficient  and pumping amplitude, respectively.
The studies of the Lugiato-Lefever equation were boosted by the  discovery of the frequency combs associated with the Kerr temporal solitons (see e.g. \cite{LL01_13,LL02_13,LL03_14}). The frequency combs are of  the utmost importance for the metrology applications \cite{Combs01,Combs02}.

Equation (\ref{LL2}) is written in its dimensionless form. The control parameters $\gamma$ and $h$ are real; the negative and positive sign in front of the cubic term corresponds to the case of normal and anomalous dispersion, respectively.
Assuming that the field is stationary and decomposing $\psi(x)=u_1(x)+iu_2(x)$,  equation (\ref{LL2}) is cast in the form
\begin{align}
&\frac12u_{1,xx}+u_1-\gamma u_2\pm\left(u_1^2+u_2^2\right)u_1-h=0,\label{LL01}\\[2mm]
&\frac12u_{2,xx}+\gamma u_1+u_2\pm\left( u_1^2+ u_2^2\right)u_2=0.\label{LL02}
\end{align}
A variety of solutions of (\ref{LL01})-(\ref{LL02}) have been reported  in literature. These include  bright and dark solitons \cite{BZB90,BS96,BSA98,BZ99,BZ11,LL03_14,PKGG16,PGKCG16} as well as  periodic structures  \cite{PGMCG14,HTS92}.
For particular values of the control parameters, there are  kinks interpolating  between different flat backgrounds \cite{PKGG16,PGKCG16}.   As a parameter is varied, a bright soliton may transform into a kink-antikink pair \cite{PGKCG16,LLKG15}, with  the transformation involving  the snaking mechanism \cite{KW06}.
Heteroclinic connections between a periodic solution and a flat background were found in the case of the anomalous dispersion  \cite{PGMCG14} and it was conjectured  \cite{PKGG16} that no such connections exist in the case of the normal dispersion. (In what follows we argue that the connections of the above type do exist in some parameter range).

{Equations \eqref{GP_S}, \eqref{Stat_Sch_Gen01}-\eqref{Stat_Sch_Gen02} and \eqref{LL2} admit a common vector formulation:}
\begin{gather}
{\bf u}_{xx}+{\bf A}(x){\bf u}-{\bf B}({\bf u},{\bf u};x){\bf u}+{\bf h}(x)=0.   \label{GenEqStat}
\end{gather}
Here ${\bf u}(x)$ is an $n$-component real vector of unknowns, ${\bf u}(x)={\rm col}~(u_1(x),\ldots,u_n(x))$;
${\bf A}(x)$ is an $n\times n$  real matrix-valued function of $x$;
${\bf B}({\bf a},{\bf b};x)$ is a diagonal $n\times n$ real matrix where the entries $B_{k,k}({\bf a},{\bf b};x)$, $k=1,\ldots,n$, are bilinear forms of $n$-component vectors ${\bf a}$ and ${\bf b}$, with the coefficients dependent on $x$;
${\bf h}(x)$ is an $n$-component real vector-function.
%}
{In this study we focus on}  {\it soliton solutions\/} of (\ref{GenEqStat}).
These are localised solutions satisfying the boundary conditions
\begin{gather}
{\bf u}(x)\to {\bf u}^- \ \mbox{as} \  x\to-\infty,\quad\quad {\bf u}(x)\to {\bf u}^+ \ \mbox{as} \  x\to+\infty,\label{BoundC}
\end{gather}
where ${\bf u}^\pm$ are constant vectors.

{
Typically, a numerical search for soliton solutions involves  an iterative algorithm  with some initial guess for the form of the soliton. Since  the system \eqref{GenEqStat}  is nonlinear,    it is not a priori obvious whether the resulting solution is unique,
and if not --- how to set up initial guesses leading to other  solutions
coexisting for the same set of parameters.
A method providing a global view of {\it all\/}
coexising solitons would therefore be of great help.
}

In this paper, we formulate a variant of the shooting method aimed to give  a comprehensive description of {the full set of}
 localised solutions {of}  the system (\ref{GenEqStat}) with the defocusing nonlinearity. The method is based on the observation  that,
under some physically meaningful assumptions about the coefficients and potential functions  of the  system  (\ref{GenEqStat}),
most of its solutions  become infinite at a finite $x$.
We formulate conditions that allow one to recognise a singular solution before it blows up. Using these conditions we estimate the distance to the singularities $X^\pm$ and seek for bounded solutions in
{the} vicinity of solutions with ``anomalously large'' $X^\pm$.
The set of bounded solutions contains, in particular,  all  solitons ---  that is, localised solutions  with the  boundary conditions \eqref{BoundC}.

The paper is organised as follows. The basics of {our}  method are outlined in Section \ref{sec:method}.
We demonstrate that the
singular behaviour is generic for equations in the class (\ref{GenEqStat}), formulate sufficient conditions for the blow-up
 and establish an upper bound for the distance to the singularity.
In Sections~\ref{sec:examples},\ref{sec:CoupledGP} and \ref{sec:LL} the method is exemplified by the  {
$\mathcal{PT}$-symmetric complex Gross-Pitaevskii equation \eqref{GP},
the  system \eqref{2_GP01}-\eqref {2_GP02}
 of two real Gross-Pitaevskii equations,    and the Lugiato-Lefever equation  \eqref{LL2}.}
Section \ref{sec:conclusion} concludes the paper with  {a}  summary and discussion.

\section{The method}
\label{sec:method}

\subsection{Singular solutions}
\label{sec:singular}

A simple example of an equation with the required properties is
\begin{gather}
u_{xx}-u^3=0.\label{Illus}
\end{gather}
This is a particular case of the Emden-Fowler equation \cite{Bellman}. It results by setting $\bu(x)\equiv u(x)$,   $\bA(x)\equiv 0$,  $\bB(u, u; x) \equiv u^2$,
and  $\bh\equiv0$ in the system (\ref{GenEqStat}) with $n=1$. The general solution of (\ref{Illus}) is a union of two bi-parametric families,
\begin{gather}
{u_a(x)=\pm\frac{\sqrt{2}A~{\rm dn}\left(A(x-\tx);2^{-1/2}\right)}{{\rm
sn}\left(A(x-\tx);2^{-1/2}\right)}}\label{Illus_a}
\end{gather}
and
\begin{gather}
u_b(x)=\pm\frac{\sqrt{2}A~{\rm
sn}\left(A(x-\tx);2^{-1/2}\right){\rm
dn}\left(A(x-\tx);2^{-1/2}\right)}{{\rm
cn}~\left(A(x-\tx);2^{-1/2}\right)},\label{Illus_b}
\end{gather}
where $A>0$,  $\tx$ is  real
and ${\rm cn}$,  ${\rm sn}$,  ${\rm dn}$ are the Jacobi elliptic functions.
An additional one-parameter family arises by sending $A\to 0$ in (\ref{Illus_a}),
\begin{gather}
u_0(x)=\pm\frac{\sqrt{2}}{x-\tilde x},\label{IllusSol}
\end{gather}
while sending $A\to 0$ in (\ref{Illus_b}) produces the trivial solution $u=0$.

Since the denominator in the expressions for $u_{a,b,0}(x)$ has zeros, none of the solutions (\ref{Illus_a}), (\ref{Illus_b}) or
(\ref{IllusSol}) is free from singularities. (The only nonsingular solution is $u(x)=0$.) The ingredient in (\ref{Illus}) that is
responsible for the blow-up of solutions, is the repulsive (defocusing) nonlinear term $-u^3$. Indeed, any
solution of the linear equation $u_{xx}=0$ or the equation with the attractive (focusing) term $+u^3$ exists
globally in the whole real axis $x\in\mathbb{R}$.

The singular solutions continue to prevail if we consider the following well-researched
 generalisation of the Emden-Fowler equation:
 \begin{gather}
u_{xx}-p(x)|u|^\beta   {\rm sgn} (u)=0.  \label{gEF}
\end{gather}
%Singular solutions are known to be typical if $p(x)>0$ whereas when $p(x)<0$,  generic solutions exists on the whole real line  \cite{KCh93}.
 %The blow up behaviour is generic also for other generalizations of the %Emden-Fowler equation, see e.g. \cite{Dulina16_1,Dulina16_2}.}
 The singular solutions are typical for the case $p(x)>0$, $\beta>1$. (See \cite{KCh93}, Theorem 20.30.)

Another generalisation is given by  the equation
\begin{gather}
u_{xx}+(\mu-V(x))u-u^3=0,\quad \mu\in\mathbb{R},            \label{Eq_u}
\end{gather}
with several classes of real $V(x)$. As in the previous examples, most of the solutions of (\ref{Eq_u}) with the initial
conditions set at some $x = x_0$, blow up  on the real line.
%This observation was used in \cite{AlfZez} to develop a computer-assisted proof approach to Eq.~(\ref{Eq_u}).
 In the case of the parabolic and infinite double-well potentials $V(x)$, the singular behavior of solutions to (\ref{Eq_u}) was exploited to classify {\it all} localised bounded modes coexisting for the given  $\mu$ \cite{AlfZez07}. In the case of the periodic potentials, the blowing up solutions are also generic \cite{AlfAvr13}. This fact gives rise to the classification of nonlinear structures in terms of bi-infinite symbolic sequences \cite{AlfKizZez17,AlfAvr13,AlfAvr14}.

\subsection{Blow-up in the system (\ref{GenEqStat}): sufficient conditions}
\label{sec:Propos}

In fact, the singular behaviour is generic even in a vector situation, with a broad class of nonlinear terms in
the system (\ref{GenEqStat}). The blow-up can be guaranteed if the   initial condition of the corresponding Cauchy problem
is simply  ``large enough". The following proposition makes this notion precise.  \medskip

{\bf Proposition 1.} Assume the following.

\begin{itemize}

\item [(i)] ${\bf A}(x)$ is a continuous matrix function defined on $x\in\mathbb{R}^+$, and
there exists a constant $\alpha_1\in\mathbb{R}$ such that for any $x\in\mathbb{R}^+$ and any ${\bf y}\in \mathbb{R}^n$,
\begin{gather}\label{CondA}
({\bf A}(x){\bf y},{\bf y})~\leq \frac{\alpha_1}2\|{\bf y}\|^2.
\end{gather}
Hereafter $\| {\bf y} \|^2= ({\bf y}, {\bf y})$.

\item [(ii)]
 Any entry $B_{k,k}({\bf y},{\bf y};x)$, $k=1,\ldots,n$, of the diagonal matrix ${\bf B}({\bf y},{\bf y};x)$ is a
positive definite quadratic form of ${\bf y}$. Moreover, $B_{k,k}({\bf y},{\bf y};x)$, $k=1,\ldots,n$,  is a continuous function of $x$, defined in $x\in\mathbb{R}^+$ for any ${\bf y}\in \mathbb{R}^n$,
 and there exists a constant $\alpha_2>0$ such that for any $x\in\mathbb{R}^+$,  any $k=1,\ldots,n$, and any ${\bf y}\in \mathbb{R}^n$,
\begin{gather}\label{CondB}
B_{k,k}({\bf y},{\bf y};x)\geq \frac{\alpha_2}2\|{\bf y}\|^2.
\end{gather}

\item [(iii)] ${\bf h}(x)$ is a continuous function of $x$ defined on $x\in\mathbb{R}^+$, and there exists a constant $H_0>0$ such that for any $x\in\mathbb{R}^+$,
\begin{gather}
\label{CondC}
\|{\bf h}(x)\|^2\leq H_0.
\end{gather}

\end{itemize}

\medskip

Then the solution of the Cauchy problem for Eq.~(\ref{GenEqStat}) with initial data ${\bf u}(0)={\bf u}_0$, ${\bf u}_x(0)={\bf u}'_0$ such that
%\textcolor{blue}{
\begin{align}
&\|{\bf u}_0\|^2>\frac1{2\alpha_2}\left(\alpha_1+1+
2\sqrt{(\alpha_1+1)^2+4\alpha_2H_0}\right),\label{Stat03}\\[2mm]
%    &\|{\bf u}_0\|^2>M,\quad M=\frac1{2\alpha_2}\left(\alpha_1+1+\sqrt{(\alpha_1+1)^2+4\alpha_2H_0}\right),
% &\|{\bf u}_0\|^2>\frac1{2\alpha_2}\left(\alpha_1+1+\sqrt{(\alpha_1+1)^2+4\alpha_2H_0}\right),\label{Stat01}\\
&({\bf u}_0,{\bf u}'_0)\geq0,
\label{Stat02}
\end{align}
%}
blows up, i.e. $\lim_{x\to \tilde x} \|{\bf u}(x)\|=+\infty$
for some $\tx \in\mathbb{R}^+$. Moreover, $0<\tilde x\leq D$ where
\begin{gather}
D=\frac{\sqrt{2}\pi}{\sqrt{2\alpha_2 \|{\bf u}_0\|^2-\alpha_1-1}}.
\label{EstCol1}
\end{gather}

\medskip

The proof of Proposition 1 is relegated to \ref{Proof}.  \medskip

Four comments are in order.\medskip

1.  Proposition 1 states that, under certain conditions on  ${\bf A}(x)$, ${\bf B}({\bf y},{\bf y};x)$ and ${\bf h}(x)$,   $x\in \mathbb{R}^+$,   the
solution of  the system (\ref{GenEqStat}) with initial conditions at $x=0$   has a singularity within the positive $D$-neighbourhood of $x=0$.
If  the assumptions (i)-(iii) are met by the coefficient matrices ${\tilde {\bf A}}(x)= {\bf A}(-x)$, ${\tilde {\bf B}}({\bf y},{\bf y};x)={\bf B}({\bf y},{\bf y};-x)$ and vector ${\tilde {\bf h}}(x)= {\bf h}(-x)$,
with $x\in\mathbb{R}^+$, a singularity is guaranteed to occur within the {\it negative\/}  $D$-neighbourhood.
%Finally, if the conditions (\ref{CondA})-(\ref{CondC}) are satisfied by  both sets, ${\bf A}(x)$, ${\bf B}(x), {\bf h}(x)$ and ${\tilde {\bf A}}(x)$, ${\tilde {\bf B}}(x), {\tilde {\bf h}}(x)$,
%with  $x\in \mathbb{R}^+$ in either case, then
%the solution %of Eq.~(\ref{GenEqStat})
%has singularities both in the positive and negative semiaxis.
\medskip

2. The initial conditions for the Cauchy problem in Proposition 1 are set at $x=0$. However, by shifting $x\to x+x^+$ the initial conditions can be imposed at any point $x=x^+$. Then Proposition 1 guarantees the presence of singularities of the solution in the positive $D$-neighbourhood of  $x=x^+$.\medskip

3. When ${\bf h}(x)\equiv 0$  the estimations of Proposition 1 can be specified. Namely, if the condition (\ref{Stat02}) holds together with
\begin{gather}
\|{\bf u}_0\|^2\geq \max\left(\frac{\alpha_1}{2\alpha_2};\frac{3\alpha_1}{2\alpha_2}\right),
\label{EstCol3}
\end{gather}
(instead of (\ref{Stat03})) then the estimation $0<\tilde x\leq D$ for the blow-up point holds with
\begin{gather}
D=\frac{\sqrt{2}\pi}{\sqrt{2\alpha_2 \|{\bf u}_0\|^2-\alpha_1}},
\label{EstCol2}
\end{gather}
instead of (\ref{EstCol1}).

%4. The estimation (\ref{EstColl}) can be used for rigorous justification %of the singularity position $x_0$. However, in numerical practice one can %make use of the observation that $\|u(x)\|^2\sim \Gamma/(x-x_0)^2$ where %$\Gamma$ is some constant and to detect the position of the singularity %comparing the behaviour of numerical solution with this asymptotics.

4.  The assumptions of Proposition 1 are satisfied by a number of physically relevant models, see the
sections \ref{sec:examples},~\ref{sec:CoupledGP} and \ref{sec:LL}.

\subsection{Numerical approach}
\label{sec:technique}

\begin{figure}%[h]
%\centerline{\includegraphics [scale=0.6]{Singular.eps}}
\centerline{\includegraphics [scale=0.8]{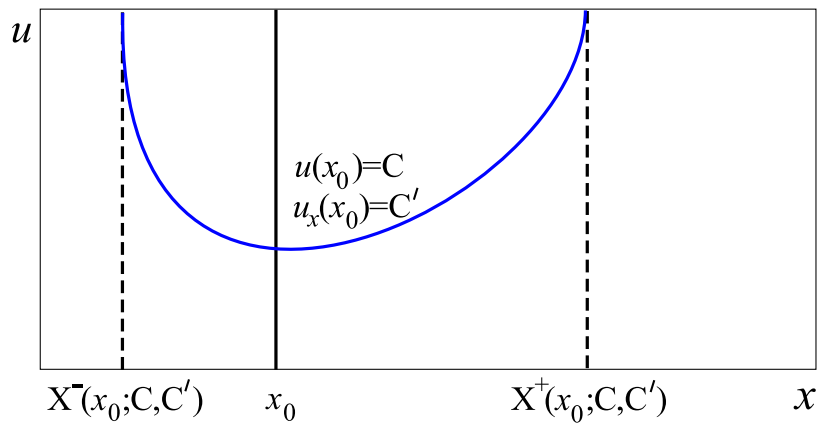}}
\caption{Definition of  the functions $X^\pm$  in the case where the system  (\ref{GenEqStat})
has a single component   ($n=1$).
 }\label{Sing_Def}
\end{figure}

Any  solution of the system  (\ref{GenEqStat}) is uniquely determined by $2n$ initial conditions imposed at some point $x=x_0$:
\begin{gather}
u_k(x_0)=C_k,\quad u_{k,x}(x_0)=C_k',    \quad k=1,\ldots,n.
\label{InDatU}
\end{gather}
Let  $X^+ (x_0; C_1, C'_1, \ldots, C_n, C'_n)$  be the coordinate of the singularity of the solution to the Cauchy problem  (\ref{GenEqStat}), (\ref{InDatU}) located {\it to the  right} of $x_0$. Similarly,
let $X^-(x_0;C_1,C'_1,\ldots,C_n,C'_n)$ be defined as the coordinate  of the singularity {\it to the left} of $x_0$.   If the solution of the Cauchy problem exists in the whole semi-infinite interval $(x_0, +\infty)$,  we write
\begin{gather}
X^+(x_0;C_1,C'_1,\ldots,C_n,C'_n)=+\infty.\label{h_inf+}
\end{gather}
Similarly, if the solution exists in the whole semiaxis $(-\infty; x_0)$, we let
\begin{gather}
X^-(x_0;C_1,C'_1,\ldots,C_n,C'_n)=-\infty.\label{h_inf-}
\end{gather}
The definition of the  functions $X^\pm$ is illustrated in Fig.~\ref{Sing_Def}.

Our aim is to detect points in the space $C_k, C_k'$ $k=1,\ldots,n$, where
equations (\ref{h_inf+}) and (\ref{h_inf-}) hold true --- that is, detect the initial conditions for which the solution  exists on the whole real line.
Note that the validity of  (\ref{h_inf+}) and (\ref{h_inf-}) does not yet imply that  ${\bf u}(x)$ is a {\it soliton} solution of Eq.~(\ref{GenEqStat}). One still needs to
check whether the boundary conditions (\ref{BoundC}) are satisfied.

In order to compute numerically $X^+$ for initial data $C_k, C_k'$ $k=1,\ldots,n$ we choose the value $\mathcal M$ large enough, such that
\begin{gather*}
\frac{\sqrt{2}\pi}{\sqrt{2\alpha_2 {\mathcal M}^2-\alpha_1-1}}<\delta x
\end{gather*}
where $\delta x$ is a small tolerance value chosen beforehand. For a given set of $C_k$ and $C_k'$, $k=1,\ldots,n$, we solve the  Cauchy problem  (\ref{GenEqStat}), (\ref{InDatU}) along the $x$-line. Let $x$ reach a point $x^+$ where the solution satisfies the conditions $({\bf u}(x^+),{\bf u}'(x^+))\geq 0$ and $\|{\bf u}(x^+)\|^2\geq{\mathcal M}^2$. Then according to Proposition 1, $x^+$ approximates the position of the singularity $X^+$ up to the error $\delta x$. In a similar manner the value of $X^-$ can be computed.
%}

Then the approach can be sketched in general as follows. Introduce a fine grid in the space of parameters $C_k,C_k'$, $k=1,\ldots,n$. For each point of the grid we compute the values $X^\pm$.  If at some node of the grid the values of $X^\pm$ are ``anomalously large'', we repeat the procedure on a finer grid in the vicinity of this node. Repeating in such a way we can locate the initial data $C_k,C_k'$, $k=1,\ldots,n$ that correspond to the bounded solution.

Evidently, the scanning in the space of parameters $C_k,C_k'$, $k=1,\ldots,n$ with large $n$ is a computationally costly exercise.
However, the dimension of the space may be reduced by noting a  symmetry of the nonlinear mode or using its asymptotic behaviour at infinity.

{\it (a) Symmetry-based reduction.} The former approach can be exemplified by the Gross-Pitaevski equation (\ref{GP_S}) with a $\PT$-symmetric potential $V^*(-x)=V(x)$. It is known that all soliton solutions supported by a generic potential of this type have to be $\PT$ symmetric \cite{KYZ16}. This implies that while studying the system (\ref{GP01})-(\ref{GP02}) one can impose the conditions $u_{1,x}(0)=0$ and $u_2(0)=0$ and analyze functions of two arguments rather than four: $X^-(0;C_1,0,0,C_2')$ and $X^+(0;C_1,0,0,C_2')$. Plotting
the function $X^+(0;C_1,0,0,C_2')$
over $(C_1,C_2')$-plane we identify
the initial values corresponding to the infinite interval of existence of the solution to the Cauchy problem. (On the other hand,
we do not need to plot the function $X^-$ since its graph obtains from the graph of $X^+$ by a mere reflection $X  \to-X$, $C_2'\to-C_2'$.)
 Once the solutions existing over the entire real line have been identified, their asymptotic behaviours as $x\to\pm\infty$ should be further examined to select solitons.
\medskip

{\it (b) Asymptotic reduction.} An alternative approach is suitable for
solutions of the system (\ref{GenEqStat}) approaching the equilibrium state ${\bf u}^+$ as $x\to+\infty$ (or the equilibrium state ${\bf u}^-$ as $x\to-\infty$). In the phase space of the dynamical system generated by equations (\ref{GenEqStat}), the corresponding trajectories lie on the stable/unstable manifold of this equilibrium state.
Accordingly, we can restrict ourselves to the evaluation of
the function $X^-$ (or $X^+$) just on that manifold. If the dimension of the stable manifold  is much smaller than $2n$, this reduction will simplify the analysis quite considerably.

To illustrate this idea, we consider the linearisation of the system (\ref{GenEqStat}) about the equilibrium state ${\bf u}^+$:
\begin{gather}
{\bf v}_{xx}={\bf L} {\bf v}.
\label{LinGenEq}
\end{gather}
Here ${\bf L}={\bf L}(x)$ is an $n \times n$ real matrix.
%
% We denote ${\bf L}^+(x)$ the dominant asymptotic behaviour of this matrix as $x \to +\infty$ and
%assume that the equation
% \begin{gather}
% {\bf v}_{xx}={\bf L}^+(x) {\bf v}
% \label{LinGenEq}
% \end{gather}
Assume that the equation (\ref{LinGenEq}) has $m$ linearly independent solutions ${\bf v}^+_1(x), ..., {\bf v}^+_m(x)$ satisfying ${\bf v}^+_i \to 0$ as $x \to +\infty$  and $m < 2n$.
 Any solution  of the vector equation (\ref{GenEqStat})  approaching ${\bf u}^+$ as $x \to \infty$, can be approximated by
\begin{eqnarray}
&&{\bf u}(x)\approx {\bf u}^+  +C_1{\bf v}^+_1 (x) +\ldots +C_m{\bf v}^+_m (x),\label{InitDataF}\\
&&{\bf u}_x(x) \approx    C_1  {\bf v}^+_{1,x}(x)    +  \ldots  + C_m {\bf v}^+_{m,x}(x).\label{InitDataDF}
\end{eqnarray}
Here $C_1,\ldots,C_m$  are real coefficients.
% ${\bf v}^+_k(x)$, ${\bf v}^+_{k,x}(x)$, $k=1,\ldots,m$ in (\ref{InitDataF})-(\ref{InitDataDF}) can be replaced by its dominant asymptotic behaviour as $x\to +\infty$.
We  examine the solution of the system (\ref{GenEqStat}) with the initial conditions  (\ref{InitDataF})-(\ref{InitDataDF})
imposed at a point $x=x_0$ with a large  $x_0>0$.
When $m=1$, one can plot  the function $X^-(x_0; C_1)$ over  an interval $-L< C_1<L$. When $m=2$, the function $X^-(x_0; C_1,C_2)$ can be plotted over a box $-L<C_1,C_2<L$.
The objective is to determine the values of the coefficients $C_1,\ldots,C_m$ giving rise to
 ``anomalously'' large negative $X^-$.
 These values correspond to large (potentially infinite)
 intervals of existence of the solution of  the Cauchy problem.
 The resulting set of solutions should then be classified into (a)  solitons and (b) nonlocalised bounded modes, such as periodic and quasiperiodic structures, flat-periodic connections etc.

\section{The $\PT$-symmetric Gross--Pitaevskii equation}
\label{sec:examples}

In this and the next two sections, our method is illustrated by
 several systems of physical significance.
We start with the  $\PT$-symmetric Gross-Pitaevskii equation.
The stationary states of the Gross-Pitaevskii model
satisfy the system (\ref{GP01})-(\ref{GP02}).
The system (\ref{GP01})-(\ref{GP02}) belongs to the general class (\ref{GenEqStat}) with ${\bf u}(x)={\rm col}~(u_1(x),u_2(x))$ and
\begin{gather*}
{\bf A}(x)=\left(
\begin{array}{cc}
\mu-V_1(x)&-V_2(x)\\
V_2(x)&\mu-V_1(x)
\end{array}\right),
\quad
{\bf B}({\bf u},{\bf u};x)=\left(
\begin{array}{cc}
u_1^2+u_2^2&0\\
0&u_1^2+u_2^2
\end{array}\right).
\end{gather*}
If $V_1(x)$ is bounded from below, the assumptions of Proposition 1 are met, with
\begin{gather*}
\alpha_1=2\left(\mu-\min_{x\in\mathbb{R}} V_1(x)\right),\quad \alpha_2=2.
\end{gather*}
According to (\ref{EstCol3}), solutions  of  the system (\ref{GP01})-(\ref{GP02}) with the initial conditions set at $x=x^+$
and satisfying
\begin{align}
&u_1^2(x^+)+u_2^2(x^+)\equiv |u(x^+)|^2\geq\frac32\left(\mu-\min_{x\in\mathbb{R}} V_1(x)\right);\\[2mm]
&u_1(x^+) u_{1,x}(x^+)+u_2(x^+) u_{2,x}(x^+)\equiv \left.\frac12\frac{d|u(x)|^2}{dx}\right|_{x=x^+}\geq 0,
\end{align}
blow up. The coordinate of the blow-up point $X^+$ is estimated as
\begin{gather}
x^+<X^+<x^+ +\frac{\sqrt{2}\pi}{\sqrt{2\alpha_2|u(x^+)|^2-\alpha_1}}.\label{Est_PT}
\end{gather}
We exemplify our approach for $\PT$-symmetric harmonic oscillator defined by
\begin{gather}
V(x)=(x-ia)^2,       \label{harmo}
\end{gather}
where $a$ is a real parameter.  In this case,
$\alpha_1=2(\mu+a^2)$. Two versions of the approach mentioned in Section \ref{sec:technique} are applicable.

\subsection{Symmetry-based approach}

The harmonic potential is generic in the sense that  it only supports  $\PT$-symmetric solutions.
These have even $u_1(x)$ and odd $u_2(x)$; consequently,
$u_{1,x}(0)=0$ and $u_2(0)=0$. The $\PT$-symmetric solutions are characterised by the initial data  $u_1(0)=C_1$ and $u_{2,x}(0)=C_2'$. (See  the discussion in Sec.~\ref{sec:technique}).

\begin{figure}%[t]
\centerline{
\includegraphics[width=1.0\columnwidth]{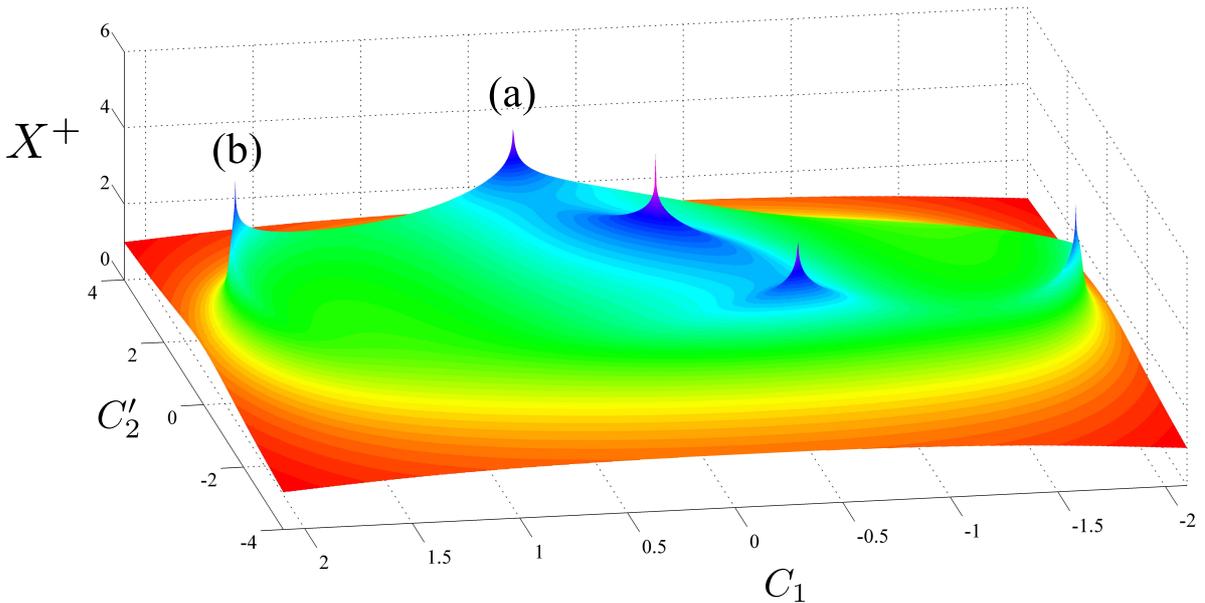}}
%\centerline{\includegraphics[width=1.0\columnwidth]{Fig2.eps}}
\caption{A numerical plot of the function $X^+(0; C_1, 0, 0,  C_2')$ for the stationary Gross-Pitaevskii equation (\ref{GP_S}) with the
$\PT$-symmetric parabolic potential $V(x)=(x-ia)^2$. In this plot,
$\mu=4$ and  $a=0.25$. The coordinates on the horizontal plane are $C_1=u_1(0)$ and $C_2'=u_{2,x}(0)$.
The bounded solutions corresponding to the peaks marked (a) and (b) are shown in Fig.~\ref{fig:x2modes}.
}\label{fig:PT3D}
%	\end{center}
\end{figure}

A numerically generated   plot of the function $X^+(0; C_1, 0, 0, C_2')$ for $a=0.25$ and   $\mu=4$ is presented in Fig.~\ref{fig:PT3D}.
In order to obtain this figure, the subset $(-2.1,2.1)\times(-4, 4)$ of the plane $(C_1,C_2')$ was sampled with   small enough increments   in $C_1$ and $C_2'$.
(Specifically, we used $\Delta C_1 = \Delta C_2'= 0.004$).
 The Cauchy problem was solved using a Runge-Kutta routine with the step size $\Delta x=0.01$.
 The numerical value of  $X^+(0; C_1, 0, 0, C_2')$ was computed by  detecting the point on the real axis  where the absolute value  of $u(x)$ exceeded a
  large enough value $\mathcal M$ chosen beforehand. Specifically, we took $\mathcal M=10^4$; in view of the bound (\ref{Est_PT})
  the resulting value of $X^+$ is accurate to within $\delta x\sim10^{-2}$. No further accuracy improvement is possible without the Runge-Kutta stepsize reduction.

In the plot shown in Fig.~\ref{fig:PT3D}, five separate peaks are clearly distinguishable. One of these corresponds to the trivial solution $u(x)\equiv 0$ while
the other four peaks correspond to  localised nonlinear modes.
The nontrivial solutions
form two pairs  symmetric with respect to the origin on the $(C_1, C_2')$-plane: $X^+(0; C_1, 0, 0, C_2') = X^+(0; -C_1, 0, 0, -C_2')$.
This symmetry stems from  the  invariance
of the equation (\ref{GP_S})  under the transformation $u \to -u$.
Accordingly, the parameter sets $(C_1, C_2')$ and $(-C_1, -C_2')$ can be regarded equivalent.

\begin{figure}%[t]
	\centerline{
		\includegraphics[width=.8\columnwidth]{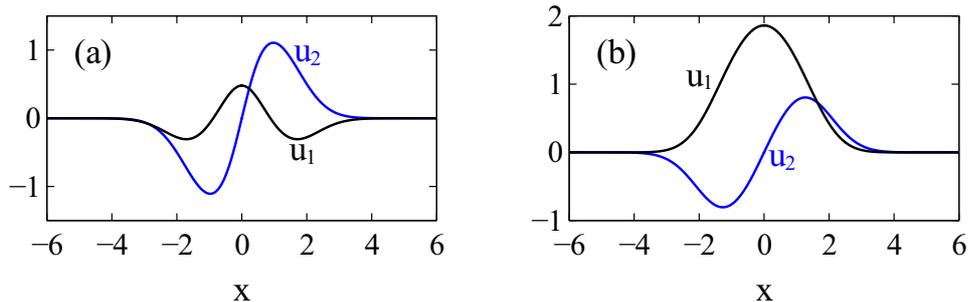}}	
		\caption{The nonlinear modes associated with two nonequivalent peaks in  Fig.~\ref{fig:PT3D}.
	The panel (a) shows the mode with $C_1=0.479$ and $C_2'= 1.972$
	and panel (b) plots the solution with  $C_1=1.860$ and $C_2'= 0.999$.
	 The black and blue curves describe $u_1(x)$ and $u_2(x)$, respectively.
	}\label{fig:x2modes}
\end{figure}

The coordinates of  a peak of $X^+(0; C_1, 0, 0, C_2')$ provide an estimate for the location of the corresponding bounded solution. Once an approximate location of the peak is known,
one can scan its neighbourhood with smaller step sizes $\Delta x$, $\Delta C_1$, $\Delta C_2'$ so as to determine the location more accurately. The
resulting values of $C_1$ and $C_2'$  are then used to compute the bounded solution. The solutions associated with two
nonequivalent peaks in Fig.~\ref{fig:PT3D} are shown in Fig.~\ref{fig:x2modes}.

\subsection{Asymptotics-based approach}

An alternative approach to localised solutions of the equation
 \eqref{GP_S} with the oscillator potential \eqref{harmo}
exploits the asymptotic behaviour of the solution rather than its symmetry.
Instead of assuming that the nonlinear mode is  $\PT$ symmetric, we require it to be localized,
\begin{equation}
\lim_{x\to \infty}u(x)=\lim_{x\to -\infty}u(x)=0.\label{local_PT}
\end{equation}

Due to boundary condition (\ref{local_PT}) the asymptotics of $u(x)$ at $x\to\pm\infty$ is determined by the linearised equation
\begin{eqnarray*}
u_{xx}+\left(\mu-(x-ia)^2\right)u=0
\end{eqnarray*}
Then the behaviour of $u(x)$ at $x\to-\infty$ is described by the asymptotic of the parabolic cylinder functions \cite{AbrStegun}
\begin{equation}
\label{eq:C}
u(x) \approx C z^{(\mu-1)/{2}} e^{-z^2/2}, \quad z=x-ia, \quad x\to -\infty,
\end{equation}
where $C$ is a constant and $a$ the parameter of the  potential \eqref{harmo}.
% Asymptotics for  $u_x(x)$ can be obtained by the  differentiation of (\ref{eq:C}) with respect to $x$.
We impose the initial condition at the point $x=x_0$, with a large negative $x_0$. The value of $u(x_0)$ is set  to  \eqref{eq:C} and the value of $u_x(x_0)$ to the dominant term in the derivative of \eqref{eq:C}:
 \begin{equation}
\label{asp}
u(x_0)=  C z_0^{(\mu-1)/{2}} \exp \left( -  \frac{z_0^2}{2} \right), \quad
u_x(x_0)=  -C z_0^{(\mu+1)/{2}} \exp \left( -  \frac{z_0^2}{2} \right), \quad
z_0=x_0-ia.
\end{equation}
In view of the U(1) symmetry of the stationary Gross-Pitaevskii equation (\ref{GP_S}), it is sufficient to consider only \textit{real  positive} $C$.  Indeed, the solution $u$ corresponding to a complex $C=|C|e^{i\phi}$ is related  to the  solution with $C$ real positive by the phase rotation  $u \to u e^{-i\phi}$.

For a generic  $C>0$, the solution with the initial conditions (\ref{asp})
blows up at some finite point $X^+=  X^+(x_0; C)$. Thus the search for  localised nonlinear modes   reduces to finding the special values of $C$ in (\ref{asp}) that produce solutions remaining bounded   on the entire line. % and $x\to\infty$.

\begin{figure}
	\centering
	\includegraphics[width=\columnwidth]{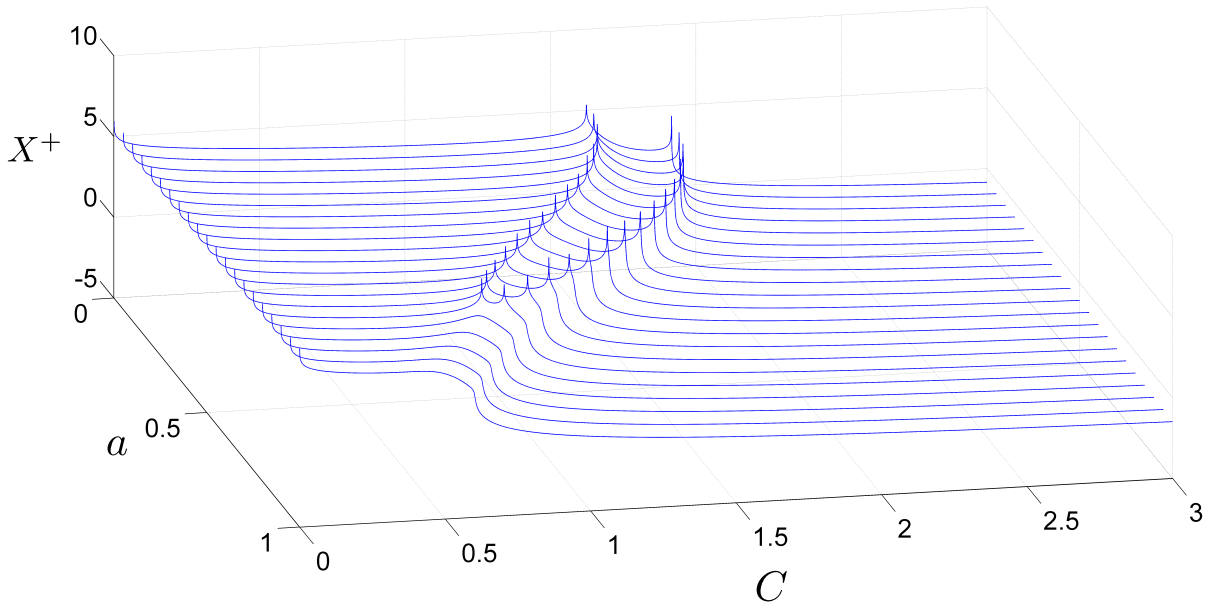}
	\caption{Plots of the function $X^+(x_0; C)$  for the stationary Gross-Pitaevskii equation (\ref{GP_S}) with $\mu=4$ and $\PT$-symmetric parabolic potential $V(x)=(x-ia)^2$.
	Here $x_0=-5$ and  $a$ ranges from $0$  to $1$.}
	\label{fig:PT2}
\end{figure}

In our numerics we used $x_0=-5$.  Figure~\ref{fig:PT2} shows the numerically generated  dependencies  $X^+(x_0; C)$
with $\mu=4$ and   twenty-one equidistant values of $a$ between 0 and 1. Each curve with  $a$ far enough from 1 features   two separate
spikes. The analysis of the shapes of the corresponding solutions $u(x)$ confirms  that each pair of  spikes  does represent
a pair of nonequivalent localised $\PT$-symmetric  modes.  (We have already determined these modes in the previous subsection
using the symmetry approach.) As  $a$ is increased, the two spikes (two  modes)   merge and disappear.

We note that these modes have been well documented in earlier literature;
see \cite{ZK12, PTGPE03}.

\section{The coupled Gross-Pitaevskii equations}\label{sec:CoupledGP}

Our second example is the system of coupled Gross-Pitaevskii equations (\ref{2_GP01})-(\ref{2_GP02}). This system also can be cast into the form (\ref{GenEqStat}) with
\begin{gather*}
{\bf A}(x)=\left(
\begin{array}{cc}
\mu_1-V(x)&0\\
0&\mu_2-V(x)
\end{array}\right),
\quad
{\bf B}({\bf u},{\bf u};x)=\left(
\begin{array}{cc}
\beta_{11}u_1^2+\beta_{12}u_2^2&0\\
0&\beta_{12}u_1^2+\beta_{22}u_2^2
\end{array}\right).
\end{gather*}
The problem that we address using our approach, can be formulated as follows.  {\it Let the  parameters  $\beta_{11},\beta_{12},\beta_{22}$ and  $\mu_{1,2}$ be assigned certain values. For this particular choice of parameters, provide a global description of
the set of all localised solutions of the system (\ref{Stat_Sch_Gen01})-(\ref{Stat_Sch_Gen02}).}

For illustrative purposes we take the harmonic  trapping potential: $V(x)=x^2$.  Normalising the coefficients  $\beta_{11}$ and $\beta_{22}$ to unity, we
assume that the remaining  parameters satisfy $\beta_{12}=\beta$, $\mu_1=\mu_2=\mu$. The system (\ref{Stat_Sch_Gen01})-(\ref{Stat_Sch_Gen02})
verifies the assumptions of Proposition 1 with $\alpha_1=2\mu$ and $\alpha_2=\max(2,2\beta)$.

Consider a class of solutions $S^-$ defined by their behaviour at the left infinity:
\begin{gather*}
 u_{1,2}(x)\to 0,~x\to-\infty.
\end{gather*}
The full asymptotic behaviour, as $x\to -\infty$,
of any solution in $S^-$  is straightforward from  the
linear part of the system (\ref{Stat_Sch_Gen01})-(\ref{Stat_Sch_Gen02}):
\begin{gather}
u_{1,2}(x)=|x|^{\frac12(\mu-1)}e^{-x^2/2}\left(C_{1,2}+o(1)\right),
\label{AsymC12}
\end{gather}
where $C_{1,2}$ are real constants. Conversely, for any choice of
  real   $C_1$ and $C_2$, there is   a unique solution of (\ref{Stat_Sch_Gen01})-(\ref{Stat_Sch_Gen02})  with the behaviour (\ref{AsymC12})
\cite{ASZ}.

%\textcolor{blue}{Note that a solution  from $S^-$ does not have to be bounded on the whole line and vanish as $x \to +\infty$. Instead, it may have singularities
%at finite $x$ and/or grow without bounds as $x \to +\infty$. }

Typically,  solutions from $S^-$ blow up at a finite point. We introduce the function
$X^+(C_1,C_2)$ that returns the point of blow up for the solution $(u_1(x),u_2(x))$  in $S^-$,
 parametrised by the pair
 $(C_1,C_2)$.
  A plot of the function $X^+(C_1,C_2)$   can be obtained by  scanning a  portion  of the plane $(C_1,C_2)$,
 with $\mu$ and $\beta$ being kept fixed.
  Since the system  (\ref{Stat_Sch_Gen01})-(\ref{Stat_Sch_Gen02}) is invariant under  $u_1(x)\to -u_1(x)$ and $u_2(x)\to
-u_2(x)$, only  the quadrant
$C_{1,2}\geq 0$ needs to be scanned.

Our numerical study proceeded by
 introducing a radial grid in the      $(C_1,C_2)$     plane.
 Having picked a large negative $x_0$,  we solved numerically
the Cauchy problem with the initial data  (\ref{AsymC12}) applied at $x=x_0$
 for each point of this grid.

According to Proposition 1 and comment 3 after this, we would terminate the numerical run
once  the condition (\ref{Stat02})
\begin{gather*}
u_1(x)u_{1,x}(x)+u_2(x)u_{2,x}(x)>0
\end{gather*}
was satisfied  and $u_1^2(x)+u_2^2(x)$ exceeded a large positive threshold ${\mathcal M}^2$,
\begin{gather*}
{\mathcal M}^2>
\max\left(\frac{\mu}{\alpha_2};\frac{3\mu}{\alpha_2}\right),
\end{gather*}
chosen beforehand.
The corresponding value of $x$, $x=x^+$,
   gives an approximation for the blow up point $X^+(C_1,C_2)$.  The error  of the approximation is bounded by equation  (\ref{EstCol2}):
\begin{gather*}
|x^+-X^+(C_1,C_2)|\leq \frac{\sqrt{2}\pi}{\sqrt{2\alpha_2{\mathcal M}^2-2\mu}}.
\end{gather*}
The values of $|x_0|$ and ${\mathcal M}$  were chosen so large that any further increase
 of these parameters did not have any visible effect on the resulting plot of  $X^+(C_1,C_2)$.

 \begin{figure}%[h]
\centerline{\includegraphics [scale=0.8]{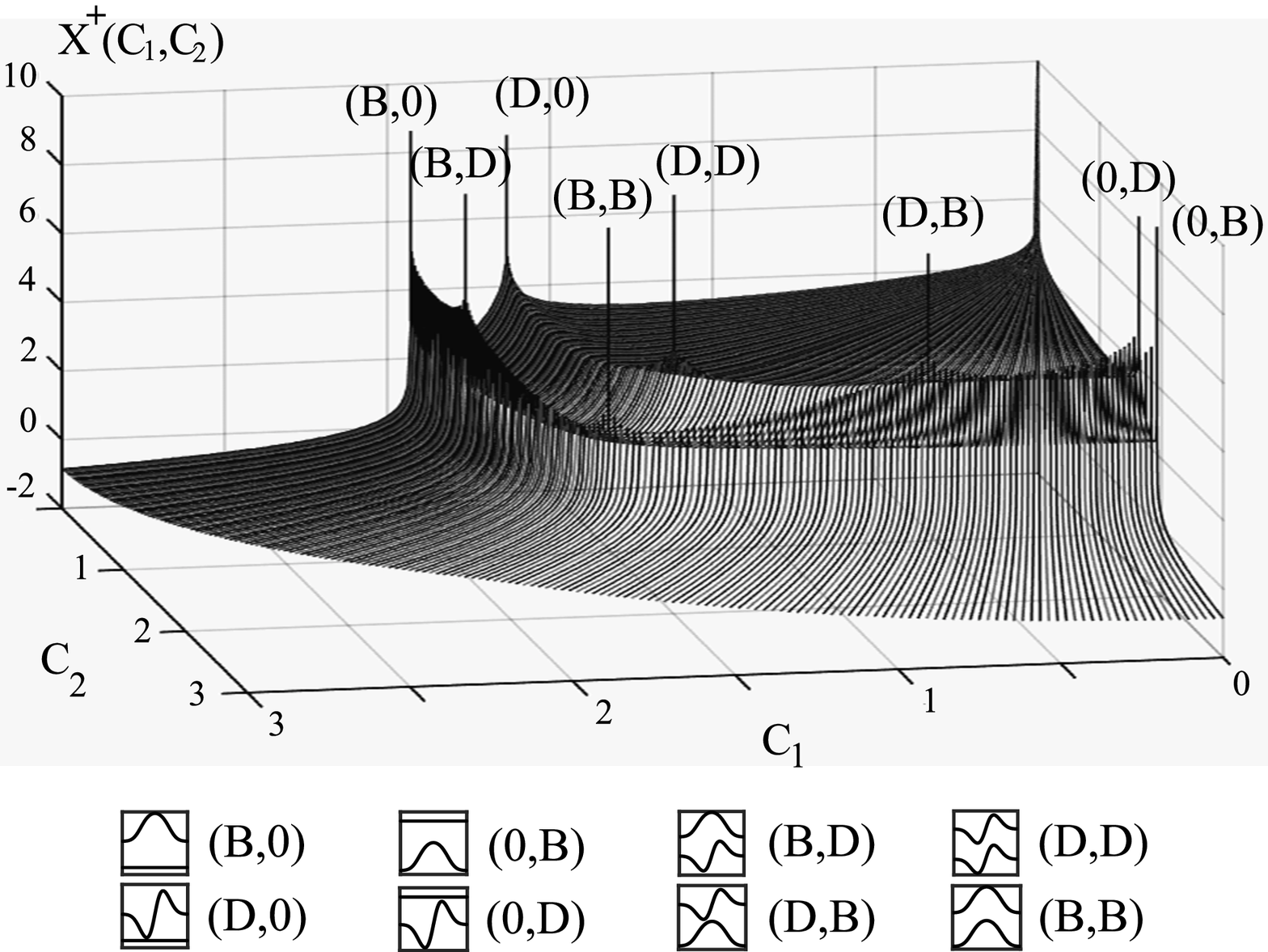}}
\caption{The function $X^+(C_1,C_2)$ for $\mu=4$ and $\beta=0.4$. The  scanned area is $C_1^2+C_2^2<9$ with $C_1\geq0$ and $C_2\geq0$.
The insets sketch the solutions corresponding to the peaks of the function. The upper and lower profiles pertain  to $u_1(x)$ and $u_2(x)$, respectively.
	%See comments in the text.
}
\label{ScanC1C2}
\end{figure}

A typical graph of the function $X^+(C_1,C_2)$ is in Fig.~\ref{ScanC1C2}. The following comments are in order. \medskip

1. The peak  at $C_1=C_2=0$   corresponds to the trivial   solution  $u_1(x)=u_2(x)\equiv 0$  of the system  (\ref{Stat_Sch_Gen01})-(\ref{Stat_Sch_Gen02}).
The two pairs of peaks on the  $C_1$ and $C_2$ axes correspond to solutions with one of the two  components equal to zero.
For instance, the peaks on the $C_1$-line correspond to solutions with $u_{2}(x)\equiv 0$ and $u_1(x)=U(x)$,  where $U(x)$ solves the equation
\begin{gather}
U_{xx}+(\mu-x^2)U-U^3=0.    \label{SingleComp}
\end{gather}
One of these solutions is  even and  nodeless  (the {\it bright} soliton), and the second one is odd and has exactly one node at $x=0$ (the {\it dark} soliton). We call these solitons {\it the single-component solitons} and mark them $(B,0)$ and $(D,0)$. The peaks on  the $C_2$-line correspond to the single-component solitons $(0,B)$ and $(0,D)$. These solutions are sketched in the insets to Fig.~\ref{ScanC1C2}.

2. The two peaks on the bisector line $C_1=C_2$ correspond to solutions with  $u_1(x)=u_2(x)=U(x)$, where $U(x)$ satisfies
\begin{gather*}
U_{xx}+(\mu-x^2)U-(1+\beta)U^3=0.
\end{gather*}
These solutions can be labelled as the {\it bright-bright} and {\it dark-dark} solitons and denoted $(B,B)$ and $(D,D)$, respectively. The components of the $(B,B)$ soliton
are even functions of $x$ and those of the $(D,D)$ soliton are odd.

3. There are two  peaks that do not belong   to any of the above-mentioned cases. These can be classified as mixed states
of {\it dark-bright} or   {\it bright-dark} type and  labelled as $(D,B)$ and $(B,D)$, respectively.

4. Refining the grid  and expanding the scanned area does not reveal any new peaks in the graph.
We accept this as numerical evidence
of  the nonexistence of any additional soliton solutions with $C_1>0$ and $C_2>0$
 for the given values of $\mu$ and $\beta$.
 The localised modes with negative $C_1$, $C_2$ or both, are related to those discussed above by reflections $u_1\to-u_1$ and $u_2\to -u_2$.
 For example, the mixed state $(B,D)$ gives rise to three solutions with negative $C_1$, $C_2$ or both. These
  will be denoted  $(-B,D)$, $(B,-D)$ and $(-B,-D)$.

\begin{figure}%[h]
\centerline{\includegraphics [scale=0.9]{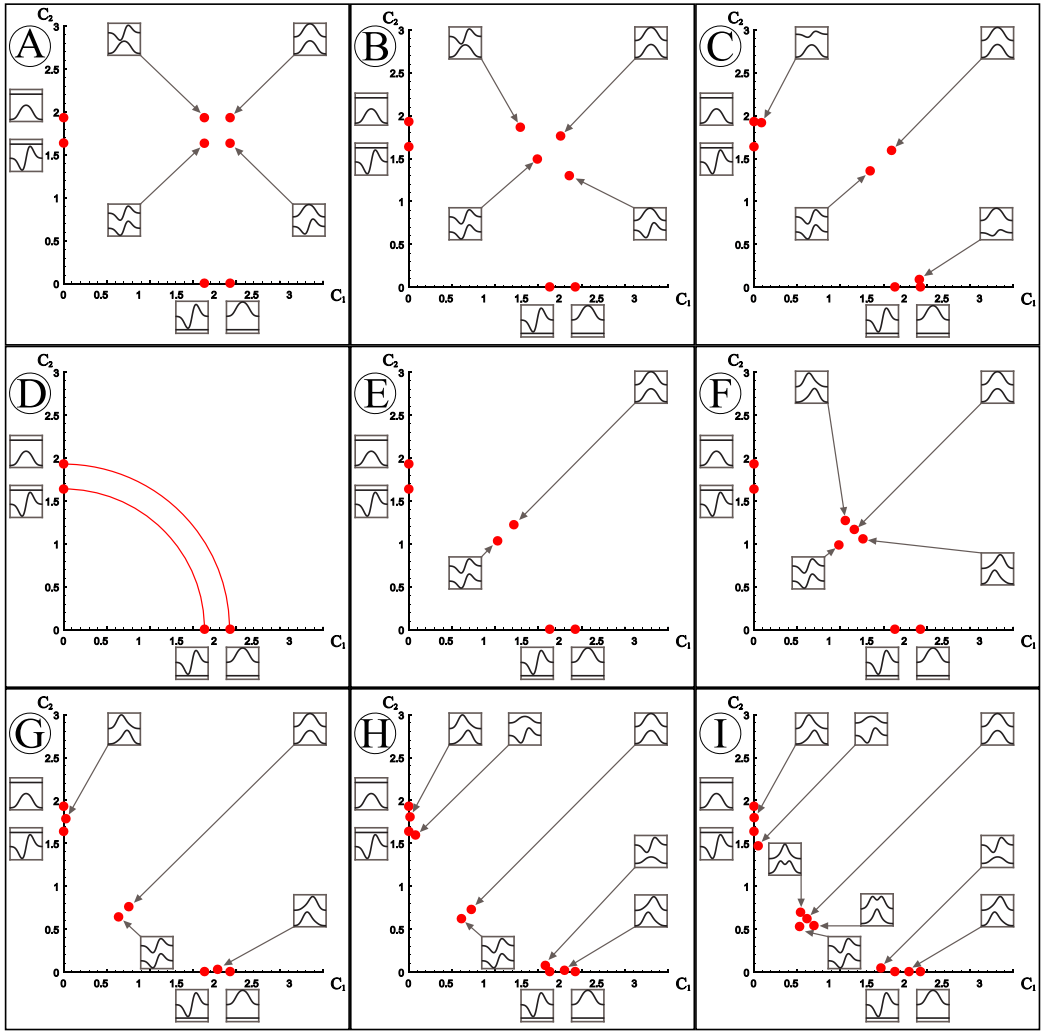}}
\caption{
Points on the parameter plane pertaining to  the soliton solutions (marked by red dots).
Here
$\mu=4$ and $0\leq \beta\leq 10$. (A) $\beta=0$; (B) $\beta=0.2$; (C) $\beta\approx \beta_1$, (D) $\beta=1$; (E) $\beta=1.5$; (F)  $\beta\approx \beta_2$; (G) $\beta=5.5$; (H) $\beta\approx \beta_3$; (I) $\beta\approx \beta_4$.
	%See comments in the text
}
\label{Bifs}
\end{figure}

Using our method, we track  bifurcations of the solitons occurring as $\beta>0$ is varied while
 $\mu$ is kept constant.
It is convenient to illustrate these bifurcations using  just the positive quadrant of the $(C_1,C_2)$    plane;  see Fig.~\ref{Bifs}. The peaks corresponding to the soliton states are marked by red dots.
As $\beta$ is changed, the solitons undergo the following transformations.

\begin{figure}%[h]
\centerline{\includegraphics [scale=0.7]{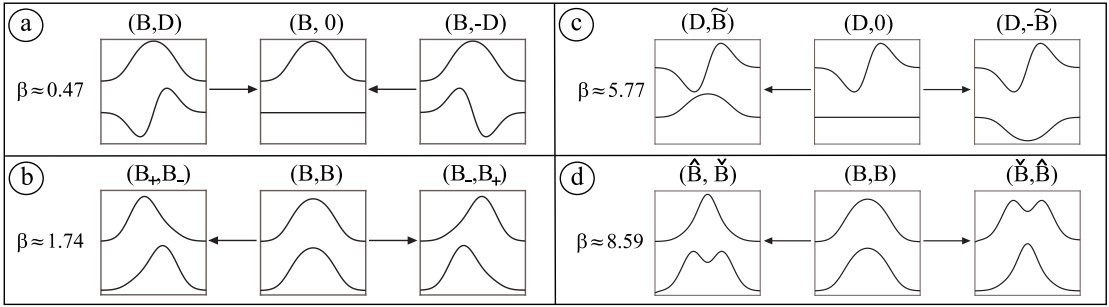}}
\caption{The schematics of the soliton bifurcations in the system (\ref{Stat_Sch_Gen01})-(\ref{Stat_Sch_Gen02})
with $\mu=4$ and $0<\beta<10$. The upper and lower profiles correspond to $u_1(x)$ and   $u_2(x)$, respectively.
}
\label{VitalBif}
\end{figure}

a) When $\beta=0$ (Fig.~\ref{Bifs}(a)) the equations  are uncoupled.  There are four nonequivalent
single-component solitons (marked by circles on the  $C_1$ and $C_2$ axes):  $(B,0)$, $(D,0)$, $(0,B)$ and $(0,D)$.  Also there are four mixed states $(B,B)$,
$(B,D)$, $(D,B)$  and $(D,D)$.
% The  $(B,B)$ and $(D,D)$ states are symmetric, i.e. $u_1(x)=u_2(x)$.

b) As $\beta$ is increased (Fig.~\ref{Bifs}(b-c)),  the points that mark  the  $(B,D)$ and $(D,B)$ states move toward the $C_1$ and $C_2$ axes, respectively.
Once the parameter $\beta$ has reached  $\beta_1\approx 0.47$,  the  solution  $(B,D)$ is absorbed by the single-component state $(B,0)$
and the $(D,B)$  is absorbed by $(0,B)$ (see the diagram \ref{VitalBif}, panel a).   Note that  the  symmetric states $(B,-D)$ and $(-D,B)$  are also absorbed by  $(B,0)$ and $(0,B)$;  therefore  the bifurcations are of the pitchfork type here.

c) Six soliton states $(B,0)$, $(D,0)$, $(0,B)$, $(0,D)$, $(B,B)$, $(D,D)$ persist in the interval $\beta_1\leq\beta<1$.
 No other  localised solutions were found in  this range of $\beta$.

d) At the point $\beta=1$, the system possesses an additional continuous symmetry.  If $(U(x),0)$ is a solution of the equation (\ref{SingleComp}), the vector function
 $(U(x)\cos\varphi,U(x)\sin\varphi)$ with $0\leq\varphi<2\pi$ satisfies the system (\ref{Stat_Sch_Gen01})-(\ref{Stat_Sch_Gen02}).
This degeneracy explains the diagram in Fig.~\ref{Bifs}(d) where the pairs of $(C_1, C_2)$ pertaining  to localised states form two circular arcs.
The arc of the smaller radius connects the $(0, D)$ and $(D, 0)$ states, whereas the longer  arc   connects  $(0, B)$ to  $(B, 0)$.

e) When $\beta$ is in the range  $1<\beta<\beta_2$, with $\beta_2 \approx 1.74$,
we have the same six soliton states again: $(B,0)$, $(D,0)$, $(0,B)$, $(0,D)$, $(B,B)$, and $(D,D)$.  [See Fig.~\ref{Bifs}(e)].
A symmetry-breaking pitchfork bifurcation occurs
as $\beta$ goes through $\beta_2$. Here,
 two asymmetric states $(B_+,B_-)$ and $(B_-,B_+)$ branch off
from the  $(B,B)$ solution [Fig.~\ref{Bifs}(f)].  The emerging states  are related to each other by the inversion $x\to -x$  supplemented by the transposition
  $u_1\to u_2$, $u_2\to u_1$ (see Fig.\ref{VitalBif}, panel b). This bifurcation corresponds to the MIT transition and has been widely discussed in the literature.

f) As $\beta$ is raised further, the $(C_1,C_2)$-points for the asymmetric states $(B_+,B_-)$ and $(B_-,B_+)$ move closer to
the  $C_1$ and $C_2$  axes  --- but never reach them. The eight soliton states
%(the six ones described in the point (c) plus two asymmetric states),
persist until the next bifurcation point $\beta=\beta_3$, with $\beta_3 \approx 5.77$ [Fig.~\ref{Bifs}(g)].

g) On passing through $\beta=\beta_3$ a new pitchfork bifurcation takes place.
Here, two pairs of dark-bright solitons with small bright components $\tilde B$
 branch off from the single-component dark solitons   $(D,0)$ and $(0,D)$ [Fig.\ref{Bifs}(h)]. The pair detaching from the  $(D,0)$ has opposite lower components;
 we denote these states $(D,\pm \tilde B)$, see Fig.\ref{VitalBif}, panel c.  The pair bifurcating from the  $(0,D)$ comprises solitons with  opposite upper components:  $(\pm \tilde B,D)$.   See Fig.~\ref{Bifs}(i).

h) One more pitchfork bifurcation occurs
 as $\beta$ passes through $\beta_4\approx8.59$.
 Here, two states branch off from the  $(B,B)$ solution. The new states
that we denote by $(\hat{B},\check{B})$ and $(\check{B},\hat{B})$, are  related by permutation: $u_1 \to u_2$, $u_2 \to u_1$. Both of their components are even and
the solutions are of the bright-bright variety, see Fig.\ref{VitalBif}, panel d.

Thus, our approach allows us to classify
 all solitons with  $\mu=4$ and track their bifurcations in the interval
$0\leq\beta\leq 10$.  The bifurcations are summarised in  Fig.~\ref{VitalBif}. It is fitting to note that the  bifurcation sequence for the $(B,B)$ solitons is well protocoled in the literature.
(See e.g. Fig.~11 in \cite{NavKevCarretero09}).
On the other hand,  the homotopy connection between the  $(B,D)$ and $(B,0)$ states (and its symmetric
  $(D,B)$ to $(0,B)$ counterpart), to the best of our knowledge, have not been reported before.
 Neither we have been not aware about the pitchfork bifurcations of  the $(D,0)$ and $(0,D)$.

\section{The Lugiato-Lefever equation}\label{sec:LL}

Our third  example is  the externally driven damped nonlinear Schr\"odinger equation --- equation (\ref{LL2}). For the stationary states $\psi =u_1(x)+iu_2(x)$,  the equation
can be written as
\begin{align}
&\frac12 u_{1,xx} +u_1   \pm   u_1 (u_1^2+u_2^2)=\gamma u_2+h, \label{St_LL01}\\
&\frac12 u_{2,xx}+u_2  \pm  u_2 (u_1^2+u_2^2)=  -\gamma u_1.  \label{St_LL02}
\end{align}
The negative and positive signs in front of the nonlinearity correspond to normal and anomalous dispersion, respectively.

In the four-dimensional  phase space with coordinates $(u_1,u_1',u_2,u_2')$ equations (\ref{St_LL01})-(\ref{St_LL02}) generate a reversible dynamical system.
(The system is conservative if $\gamma=0$; otherwise, to the best our knowledge, it does not have any
first integral.)
%(The system  is only conservative if $\gamma \neq 0$.)
Homogeneous solutions  --- often referred to as the flat backgrounds ---
correspond to equilibria of the dynamical system,
and periodic solutions  correspond to its closed orbits.
Solutions that are asymptotic to the same respectively different backgrounds as $x \to +\infty$ and $x \to -\infty$,
represent homoclinic respectively heteroclinic orbits.
There is a great body of work devoted to the closed orbits as well as  homoclinic and heteroclinic solutions in reversible systems, see \cite{D77,Ch98,H98,KW06}.    Methods for the determination of such trajectories are also well documented \cite{ChSp93}.

  %
  %
   %        In the case of the normal dispersion there are {\it platicons\/}  consisting of flat intervals connected by  kinks \cite{LLKG15}.

In the case of the normal dispersion, the system
\begin{align}
&\frac12   u_{1,xx}  +u_1-u_1(u_1^2+u_2^2) =\gamma u_2+h, \label{St_LL01_r}   \\
&\frac12     u_{2,xx} +u_2 - u_2 (u_1^2+u_2^2) =-\gamma u_1     \label{St_LL02_r}
\end{align}
satisfies the assumptions of Proposition 1 of section \ref{sec:Propos}. In this case
$\alpha_1=\alpha_2=4$, and $H_0=4h^2$, so that if
\begin{gather*}
|u(x^-)|^2\geq\frac18\left(5+2\sqrt{25+64h^2}\right);\quad \left.\frac{d|u(x)|^2}{dx}\right|_{x=x^-}\geq 0,
\end{gather*}
the solution of the Cauchy problem with the initial conditions at $x=x^-$ blows up at some point within the left $D$-neighbourhood of $x=x^-$ where
\begin{gather*}
D=\frac{\sqrt{2}\pi}{\sqrt{8 |u(x^-)|^2-5}}.
\end{gather*}

The homogeneous solutions of (\ref{St_LL01_r})-(\ref{St_LL02_r}) are $u_1(x)\equiv U_1$, $u_2(x)\equiv U_2$ where the constants  $U_{1,2}$ satisfy the algebraic system
\begin{align}
&U_1- U_1(U_1^2+U_2^2)=\gamma U_2+h \label{St_d_LL01}\\
&U_2 -U_2(U_1^2+U_2^2)=-\gamma U_1.\label{St_d_LL02}
\end{align}
Depending on the values of $\gamma$ and $h$, the system (\ref{St_d_LL01})-(\ref{St_d_LL02}) may have  one or three roots \cite{BS96,BSA98}.
The type of the equilibrium is determined by the eigenvalues $\Lambda_{1,2}$ of the           linearisation            matrix
\begin{gather*}
{\bf L}=\left(
\begin{array}{cc}
L_{11}&L_{12}\\[2mm]
L_{21}& L_{22}
\end{array}
\right),
\end{gather*}
where
\begin{gather*}
L_{11}=6U_1^2+2U_2^2-2,\quad L_{12}=2\gamma+4U_1U_2;\\[2mm]
L_{21}=-2\gamma+4U_1U_2,\quad L_{22}=6U_2^2+2U_1^2-2.
\end{gather*}

\begin{figure}%[h]
\centerline{\includegraphics [scale=0.8]{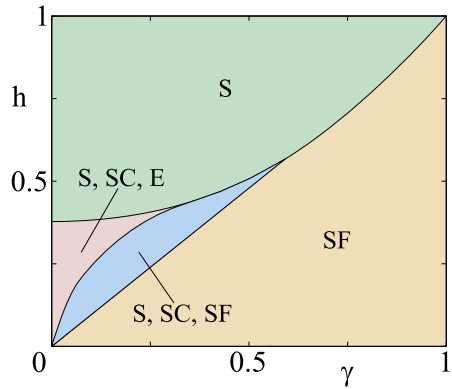}}
\caption{Four types of equilibrium states on the plane of control parameters. There is a single equilibrium (single flat background) in the green- and
 peach-tinted regions.
The equilibrium is a saddle  (marked $S$)  in the green domain and saddle-focus ($SF$) in the peach-coloured area. In each of the pink and blue domains,
 there are three coexisting flat backgrounds. The blue region harbours a saddle ($S$), saddle-center ($SC$) and
saddle-focus ($SF$) while in the pink area, the three equilibria are a saddle, saddle-center and elliptic point (marked $E$).}
\label{Zones}
\end{figure}

Since the dynamical system (\ref{St_LL01})-(\ref{St_LL02}) is reversible, it may have four types of equilibria.
(i) If the eigenvalues are both real
and positive,  the equilibrium is classified as a {\it saddle}.  (ii) Real negative $\Lambda_{1,2}$
corespond to  an {\it elliptic point}. (iii) Two real eigenvalues of the opposite sign define a {\it  saddle-center}.   (iv) Finally, an equilibrium with the complex conjugate eigenvalues, is a {\it saddle-focus}.
Fig.~\ref{Zones} divides the  $(\gamma,h)$-plane into four parameter regions according to the number of
equilibria and their types.

Generic localised solutions of (\ref{St_LL01})-(\ref{St_LL02}) are asymptotic to the flat backgrounds of  the saddle and saddle-focus    types.
To illustrate our method, we restrict ourselves to  the case of the saddle-focus
and denote $\Lambda, \overline{\Lambda}$ the associated pair of complex eigenvalues of $\bf{L}$.

 In the vicinity of the saddle-focus equilibrium, there exist a local 2D stable manifold, $W^s$, and a  local 2D unstable manifold, $W^u$.
 We parametrise trajectories on $W^s$ as follows. Let ${\rm Im}~\Lambda>0$ and define  $\lambda=\alpha+i\beta$,  with $\alpha,\beta>0$, such that
 $\Lambda=\lambda^2$.
The asymptotic behaviour of a solution of (\ref{St_LL01})-(\ref{St_LL02}) that tends to $(U_1;U_2)$ as $x\to +\infty$ is
\begin{align}
u_1(x)-U_1&\approx    - \epsilon  L_{12}e^{-\alpha x}\cos(\beta (x-\varphi));\label{AsLL01}\\[2mm]
u_2(x)-U_2&\approx   \epsilon  |L_{11}-\Lambda|e^{-\alpha x}\cos(\beta( x-\varphi)-\psi).  \label{AsLL02}
\end{align}
Here $\psi={\rm arg}~(L_{11}-\Lambda)$ while
 $0\leq\varphi<2\pi$ and  $\epsilon>0$    are free parameters.

\begin{figure}%[h]
%\centerline{\includegraphics [scale=0.6]{Method02.eps}}
\centerline{\includegraphics [scale=0.6]{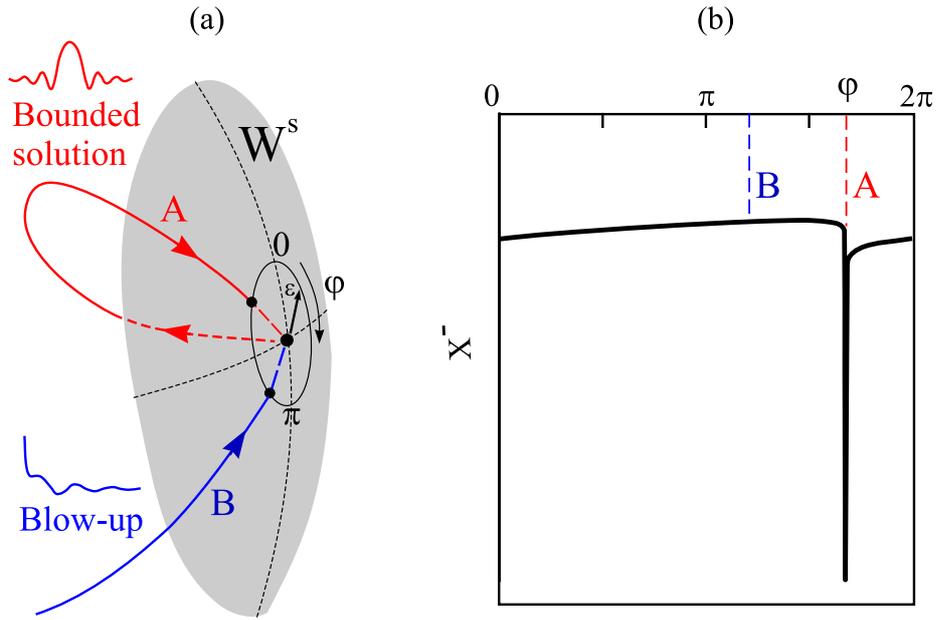}}
\caption{The left panel illustrates the stable    manifold $W^s$ of the equilibrium (shaded).
The trajectories that flow into the equilibrium as $x \to \infty$ are parametrized by the angle $\varphi$, $0\leq \varphi<2\pi$.
The right panel sketches the position of the singularity of the trajectory with the parameter $\varphi$
in the negative-$x$ line.
The function $X^-(\varphi)$
was evaluated  by solving the initial-value problem numerically.
 }\label{Method}
\end{figure}

Consider the Cauchy problem for the system (\ref{St_LL01})-(\ref{St_LL02}) with the initial data on $W^s$:
\begin{align}
&u_1(0)=U_1-\epsilon L_{12}\cos\beta\varphi;\label{Init01}\\
&u_{1,x}(0)=\epsilon L_{12}|\lambda|\cos(\beta\varphi+\theta);\label{Init02}\\
&u_2(0)=U_2+\epsilon |L_{11}-\Lambda|\cos(\beta\varphi+\psi);\label{Init03}\\ &u_{2,x}(0)=-\epsilon |L_{11}-\Lambda||\lambda|\cos(\beta\varphi+\psi+\theta),
\label{Init04}
\end{align}
%\textcolor{red}{I suggest to replace $u_{1,x}$ with $\frac{d u_1}{dx}$  for the uniformity.}
where $\theta={\rm arg}~\lambda$ and $\epsilon$ is  a fixed positive value that  has to be taken small enough.
In \eqref{Init01}-\eqref{Init04} we have used
 the translation
invariance to set the initial condition at $x=0$. Since $\epsilon$ is fixed, the parameter
$\varphi$  defines the trajectory on $W^s$ uniquely.  See Fig.~\ref{Method}.

Let $X^-$ denote  the coordinate
of the singularity of the solution with the asymptotic behaviour (\ref{AsLL01})-(\ref{AsLL02}) as $x\to+\infty$. Clearly, $X^-$ depends on the parameter $\varphi$.
(See the schematic in Fig.~\ref{Method}.)
The graph of  the function $X^-(\varphi)$ for $0\leq\varphi<2\pi$ is constructed by solving the system  (\ref{St_LL01})-(\ref{St_LL02}) in negative $x$,
with the initial data  \eqref{Init01}-\eqref{Init04}.
%\tr{ the location of the singularity can be determined by means of the bound (\ref{EstColl}).}
The value of $X^-$  is approximated by the coordinate of the point on the real
axis where $|u(x)|$ exceeds a threshold  $\mathcal M$ set
beforehand. The accuracy of this approximation is given by $D$ in equation \eqref{EstCol1}.

We now present results of the computations for $h=0.2$ and three values of $\gamma$.

\subsection{The case $h=0.2$, $\gamma=0.1$.}\label{sub01}

\begin{figure}%[h]
	\centerline{\includegraphics [scale=0.6]{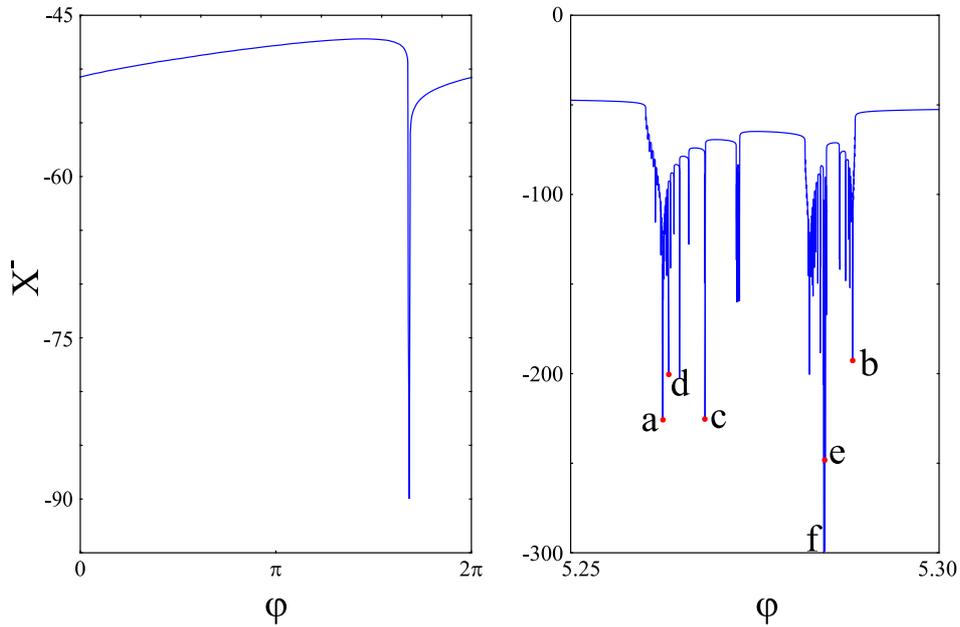}}
	\caption{
	 The function $X^-(\varphi)$ for $h=0.2$ and $\gamma=0.1$.
	In the left panel, the interval $0 \leq \varphi < 2 \pi$ is sampled
	at a low resolution.   The right panel zooms in on the neighbourhood  of the well in the left panel.
	Solutions corresponding to the  dips marked by letters  are shown in Fig \ref{Profiles}.
	The only  solution that does not blow up in our computation corresponds to the dip marked $f$.
		}\label{h_-}
\end{figure}

\begin{figure}%[h]
	\centerline{\includegraphics [scale=0.5]{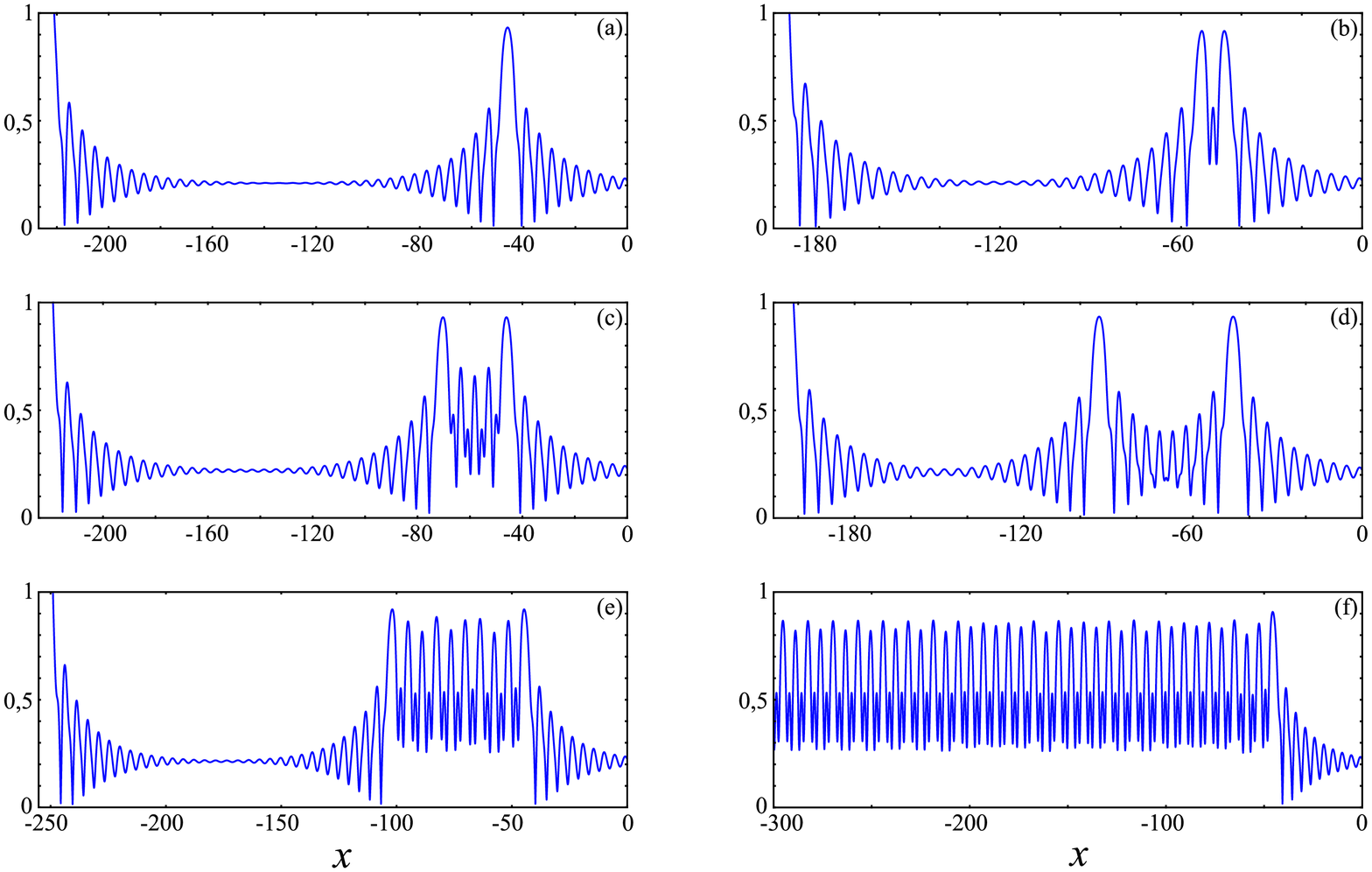}}
	\caption{Solutions of the system (\ref{St_LL01_r})-(\ref{St_LL02_r}) corresponding to the ``dips'' of the graph of $X^-(\varphi)$  in Fig.~\ref{h_-}(b).
	The panel (a) displays the solution corresponding to the dip marked $a$ in Fig.~\ref{h_-}(b);
	the panel (b) shows the solution marked $b$, and so on.
	Plotted is $|u|=\sqrt{u_1^2+u_2^2}$.
	The computations were run until $|u|$ reached $10^{4}$.
	The solution in (a) is the fundamental soliton; panels (b)-(d) show
		bound states of  fundamental solitons. Panel  (f) depicts a heteroclinic connection between the saddle-focus equilibrium
		and  a periodic orbit while panel (e) features a solution consisting of two such connections.
			}\label{Profiles}
\end{figure}

The function $X^-(\varphi)$ is depicted in Fig.~\ref{h_-}.
A coarse sampling of  the interval $0 \leq \varphi < 2 \pi$   reveals a single well;
see panel (a).
Panel (b) zooms in on the fine structure of that well.
Here, one can discern numerous dips representing bounded solutions.
Some of these solutions are illustrated in Fig.~\ref{Profiles}.

Several remarks are in order here.

1. Figure \ref{h_-}(b) gives only a rough idea
of the fine structure of the well that is extremely complex. According to \cite{H98}, the existence of the primary homoclinic orbit implies  the existence of infinitely many homoclinic orbits that make $n \geq 2$ loops in the vicinity of the
primary orbit. These orbits correspond to bound states of solitons located some distance away from each other. This means that there are infinitely many dips that should
be discernible by  a sufficiently dense sampling.

2.   Fig.~\ref{Profiles} displays  the Cauchy-problem  solutions   corresponding to some  of the dips in Fig.~\ref{h_-}(b).
Most of these solutions blow up  a certain distance away from the initial point. However this distance is large; hence
Fig.~\ref{Profiles} gives a fairly accurate description of the bounded solutions that remain close to the blowing-up solution over a long  $x$ interval.
(The former can be determined by the fine-tuning of $\varphi$; see remark 4 below.)
 In particular, panel (a) of Fig.~\ref{Profiles} gives an idea of the  {\it fundamental} soliton
of Eqs. (\ref{St_LL01_r})-(\ref{St_LL02_r}). Panels (b)-(d) show
a bound state of two fundamental solitons, with varied separation. (The existence of the bound states is due to the general argument in \cite{H98}; these solutions of the
Lugiato-Lefever  equation with normal dispersion have been reported in \cite{PKGG16,PGKCG16}).
Another type of localised mode ---  more precisely, a  solution that remains close to such a mode ---  appears in panel (e). This
solution deserves a special comment; see remark 5 below.

3. Panel (f) displays a {\it heteroclinic connection\/} between the saddle-focus equilibrium and a periodic orbit.
This solution is novel.
Although periodic solutions of the Lugiato-Lefever equation were reported in the literature \cite{HTS92}, the  connections between the flat and periodic
solutions were  only  found in the  equation with anomalous
dispersion \cite{PGMCG14}. (In fact the heteroclinic connections of the said type were conjectured not to exist in the case of the
normal dispersion \cite{PKGG16}.)

  4.
 Even though generic solutions of the initial-value problem   \eqref{St_LL01}-\eqref{St_LL02}, \eqref{Init01}-\eqref{Init04}
are  unbounded, this one-parameter family
 does contain solitons and their complexes. The corresponding special parameter values  can be determined in an algorithmic way.

Indeed, since the system is reversible,  all trajectories that pass through the symmetry plane $S=\{u_1'=u_2'=0\}$ are symmetric. This means that the solution $(u_1(x),u_2(x))$ satisfying
\begin{gather*}
u_{1}(x)\to U_1,\quad u_{2}(x)\to U_2,\quad   \mbox{as} \ x\to+\infty
\end{gather*}
and
\begin{gather*}
u_{1,x}(x^*)=u_{2,x}(x^*)=0
\end{gather*}
for some $x=x^*$, has to satisfy
\begin{gather*}
u_{1}(x)\to U_1,\quad u_{2}(x)\to U_2,\quad  \mbox{as} \ x\to-\infty
\end{gather*}
as well. This observation suggests the following {\it symmetrisation\/} algorithm.

\begin{figure}%[h]
 \centerline{\includegraphics [scale=0.5]{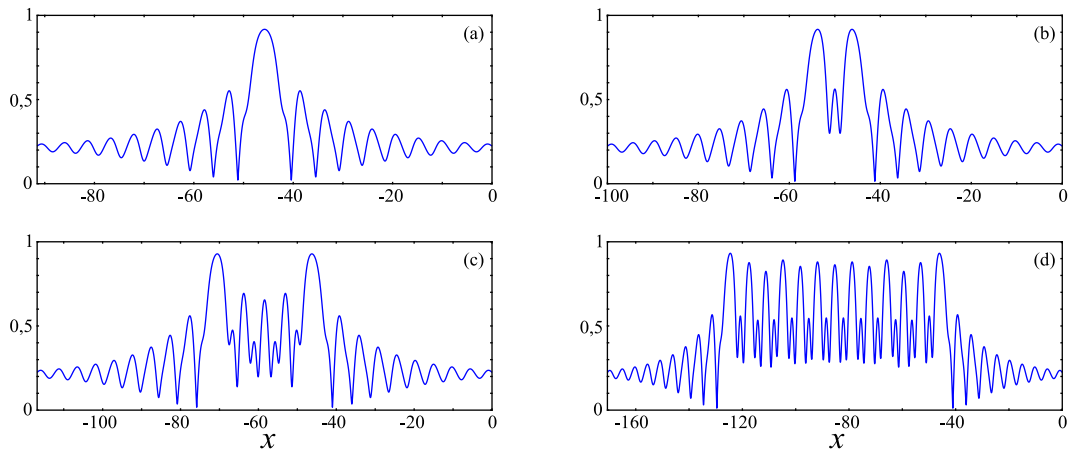}}
\caption{Soliton solutions of the system (\ref{St_LL01_r})-(\ref{St_LL02_r}) computed by means of the symmetrisation procedure.
Here   $h=0.2$, $\gamma=0.1$; the displayed quantity is
 $|u|=\sqrt{u_1^2+u_2^2}$.
  (a): the fundamental soliton;  (b)-(c): two-soliton complexes with different intersoliton separation distances;
 (d):  a localised  state with ``locked'' oscillations formed by two flat-periodic connections.
}\label{Profiles_mod}
\end{figure}

Let $\varphi$ be fixed and $u_{1}(0)$, $u_{1,x}(0)$, $u_{2}(0)$, $u_{2,x}(0)$ given by (\ref{Init01})-(\ref{Init04}).
We order zeros $x_n^*$ of the function $u_{1,x}(x)$,  $n=0,1,\ldots$,  so that
\begin{gather*}
\ldots  <x_3^*          <x_2^*<x_1^*<0
\end{gather*}
and define
\begin{gather*}
W_n(\varphi)=u_{2,x}(x^*_n),\quad n=1,2,\ldots
\end{gather*}
Roots of the function $W_n(\varphi)$ lying in the neighbourhood of a ``dip'' of the function $X^-(\varphi)$
correspond to soliton solutions of the Lugiato-Lefever equation (\ref{St_LL01_r})-(\ref{St_LL02_r}). These roots can be determined by the bisection
method. Some profiles of soliton solutions obtained by this symmetrisation procedure are displayed
in Fig.~\ref{Profiles_mod}.

5. The heteroclinic connection between a periodic solution
 and  the saddle-focus equilibrium gives rise to novel localised states with ``locked'' oscillations.  A solution of this type is shown in Fig.~\ref{Profiles_mod} (d).
This solution has been obtained by the symmetrisation procedure in the neighbourhood of the heteroclinic connection.
 (The above structure bears some similarity  to the ``truncated Bloch waves'' of the  Gross-Pitaevskii equation with a periodic potential \cite{WYAK09}.)

\subsection{The case $h=0.2$, $\gamma=0.17$.}\label{sub02}

\begin{figure}%[h]
\centerline{\includegraphics [scale=0.5]{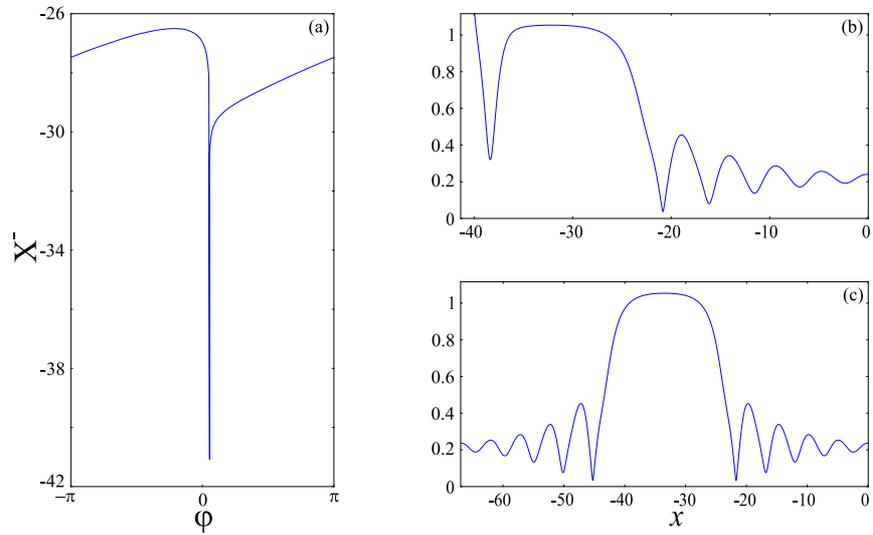}}
\caption{(a): The function $X^-(\varphi)$ for $h=0.2$ and $\gamma=0.17$. (b): The solution corresponding to the lowest value of $X^-(\varphi)$
in panel (a) that we were able to reach.
(c): A soliton obtained  by means of the symmetrisation procedure in the neighbourhood  of the solution in panel (b).
 }\label{LL_Exam02}
\end{figure}

The function $X^-(\varphi)$ with $h=0.2$, $\gamma=0.17$ has just one narrow dip; see
Fig.~\ref{LL_Exam02} (a).
 Our numerical resolution was insufficient to discern any internal structure of this dip.
 The initial-value problem solution with the largest negative value of $X^-(\varphi)$ that we were able to reach, is shown in Fig. \ref{LL_Exam02} (b).
 Making use of  the symmetrisation algorithm in the neighbourhood of this solution we construct a localised nonlinear mode (Fig. \ref{LL_Exam02} (c)).
 This soliton can be interpreted as  a
 kink-antikink pair, where each of the kink and antikink  connect the saddle and  saddle-focus equilibria.
 Localised solutions of this type as well as the transition from the fundamental soliton to the kink-antikink pair by means of the ``snaking'' scenario, were discussed in \cite{PGKCG16}.

\subsection{The case $h=0.2$, $\gamma=0.4$.}\label{sub03}
\begin{figure}%[h]
\centerline{\includegraphics [scale=0.4]{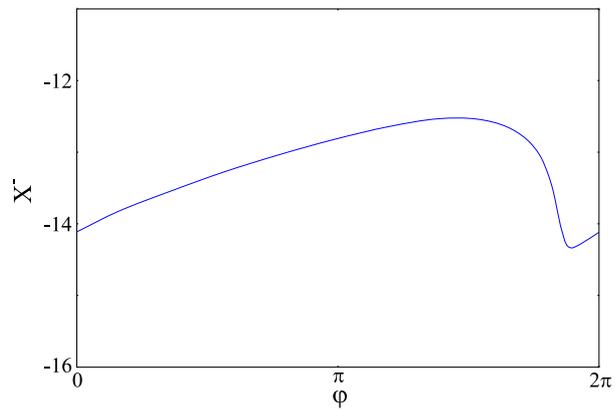}}
\caption{The  function $X^-(\varphi)$ for  $h=0.2$ and $\gamma=0.4$.
}\label{LL_Exam03}
\end{figure}

The  function $X^-(\varphi)$ with $h=0.2$, $\gamma=0.4$ is shown in Fig.~\ref{LL_Exam03}.
As it does not feature any dips,  we conclude that no bounded solutions exist in this case.  As $\gamma$ is increased,
the function $X^-(\varphi)$ flattens out so that
for large  $\gamma$ it is nearly a constant.
This observation suggests that  no soliton solutions asymptotic to the saddle-focus equilibrium exist for large values of $\gamma$.

A more detailed study of the novel nonlinear modes of the Lugiato-Lefever equation will be presented in a forthcoming publication.

\section{Concluding remarks}
\label{sec:conclusion}

In this paper, we have described a method for the numerical search and computation of localised solutions to a family       of
 scalar or vector Schr\"odinger-type
  equations with the defocusing nonlinearity.
  Our method makes use of the fact that
  most of the solutions to the system (\ref{GenEqStat}) blow up a finite distance away from the initial point on the real line.
  It is, effectively, a procedure for  the filtering out of the singular solutions.
The procedure
relies on a set of sufficient
conditions for the blow up and  an expression for an upper bound on the distance to the singularity. The sufficient conditions and {the} upper bound are derived in this paper.

Unlike  iterative methods, our approach does not require any a priori details on the shape of  solution that we seek to determine.
As a result, the method is {\it global\/}: by scanning a sufficiently large domain in the space of initial values, we can {obtain}  a comprehensive description of the set of coexisting localised modes.
Its additional advantage  is that it is capable of establishing the {\it nonexistence\/}  of regular solutions
to the boundary-value problem in question. If the nonlinear system does not have any localised modes,
 the method yields a ``numerical evidence'' of this fact.

We have illustrated our method with
three examples. First, we used it to reproduce several results from literature on
localised modes in the nonlinear Schr\"odinger equation with a complex-valued $\PT$-symmetric potential.
Second, we applied the new approach to a system of coupled Gross-Pitaevskii equations. Here, we have determined  all coexisting soliton solutions in a chosen parameter
range  and tracked their bifurcations as the coupling parameter is varied. The third example concerned  the Lugiato-Lefever equation
with normal dispersion.
In the latter case the method produces some novel solutions in addition to structures described  in the literature.
The new solutions include the  heteroclinic connection between a periodic orbit and equilibrium state
as well as  a sequence of localised solutions consisting of several periods of oscillation connected to the equilibrium state on either side.

We close the paper by mentioning several theoretical and practical  aspects of the method that deserve further investigation.

First, it would be instructive to fully understand  properties of the functions $X^{\pm}$ that play the central role in the elimination of
singular solutions.
Under what  conditions are these functions (a) well  defined, (b) continuous, (c) differentiable, and (d) monotonic?  Some results for particular cases can be found in
\cite{KCh93,Dulina16_1,Dulina16_2,Korchem16,AlfKiz16}.

Second, one should try to loosen
the conditions (\ref{CondB}) on the nonlinear term that guarantee the formation of  a singularity. The result of \cite{LiHuang02} may be used
to generalise the conditions (\ref{CondB})   { beyond the cubic} defocusing nonlinearities. {On the other hand,} equations that have neither focusing nor defocusing nonlinearity but exhibit blowing up are not unheard of in the literature.
For instance, most of the solutions to  the equation
\begin{gather}
u_{xx}+\mu u-P(x)u^3=0,\label{NonAutNon}
\end{gather}
where $P(x)$ is  a real function with alternating sign, develop a singularity at a finite point in $\mathbb{R}$ \cite{AlfLeb,AlfLebMal}.
With regard to the system (\ref{2_GP01})-(\ref{2_GP02}), this suggests
  that our method may remain effective even for   {a}  much wider class of nonlinearities.

Finally, the present formulation of our approach confines it to the one-dimensional geometry. The method does not admit a
straightforward generalisation to  systems of the form  (\ref{GenEqStat}) with
 $\bu_{xx}$  replaced with   $\Delta \bu$.
However it may be possible to extend it to  quasi-one-dimensional situations,
in particular to the radially-symmetric solutions in two and three dimensions.

\section*{Acknowledgments}

Authors are grateful to A.D. Kirilin for {his help with}  numerical work and fruitful discussions. GA and DZ were funded by  Russian Science Foundation (Grant No. 17-11-01004).
 IB was  supported by the National Research
Foundation of South Africa (grants 105835, 85751 and 466082)
and
the European Union's Horizon 2020 research
and innovation programme under the Marie Sk{\l}odowska-Curie
Grant Agreement No. 691011.

\appendix

\section{Proof of Proposition 1 in Section \ref{sec:singular}}\label{Proof}

Having taken  the dot product of equation (\ref{GenEqStat}) with ${\bf u}$,
\begin{gather*}
({\bf u}_{xx},{\bf u})+({\bf A}(x){\bf u},{\bf u})-({\bf B}({\bf u},{\bf u};x){\bf u},{\bf u})+({\bf h}(x),{\bf u})=0,
\end{gather*}
we make use of (\ref{CondA})--(\ref{CondC}) and the inequalities
\begin{align*}
&({\bf u}_{xx},{\bf u})=\frac12 ({\bf u},{\bf u})_{xx}-({\bf u}_x,{\bf u}_x)\leq \frac12(\|{\bf u}\|^2)_{xx}, \\[2mm]
&({\bf h}(x),{\bf u})\leq\frac12\left(\|{\bf h}(x)\|^2+\|{\bf u}\|^2\right),
\end{align*}
%This gives the following relation:
%\begin{gather*}
%(\|{\bf u}\|^2)_{xx}+\left(\alpha_1+1\right)\|{\bf u}\|^2-\alpha_2\|{\bf u}\|^4+\|{\bf h}(x)\|^2\geq 0,
%\end{gather*}
%i.e.
to obtain
\begin{gather}\label{DiffIn}
(\|{\bf u}\|^2)_{xx}\geq-\left(\alpha_1+1\right)\|{\bf u}\|^2+\alpha_2\|{\bf u}\|^4-H_0.
\end{gather}
Consider an auxiliary scalar equation
\begin{gather}\label{DiffEqV}
v_{xx}=-(\alpha_1+1)v+\alpha_2v^2-H_0.
\end{gather}
Eq.~(\ref{DiffEqV}) has two homogeneous solutions:
\begin{align*}
&\tilde v_1=
\frac1{2\alpha_2}\left(\alpha_1+1-\sqrt{(\alpha_1+1)^2+4\alpha_2H_0}\right)\leq0,\\[2mm]
&\tilde v_2=
\frac1{2\alpha_2}\left(\alpha_1+1+\sqrt{(\alpha_1+1)^2+4\alpha_2H_0}\right)\geq0.
\end{align*}
The solution of (\ref{DiffEqV}) with initial data $v(0)=v_0$, $v_x(0)=v'_0\geq 0$ can be written in the implicit form
\begin{gather}\label{ImplSolG}
x=\int_{v_0}^v
\frac{\sqrt{3}d\xi}
{\sqrt{2\alpha_2(\xi^3-v_0^3)-
3(\alpha_1+1)(\xi^2-v^2_0)-6H_0(\xi-v_0)+3(v'_0)^2}}.
\end{gather}
When $v_0>\tilde v_2$,
the denominator in (\ref{ImplSolG}) has no roots between $v_0$ and infinity. Then, the integral (\ref{ImplSolG}) converges for $v=\infty$. This implies that any solution of (\ref{DiffEqV}) with initial conditions $v(0)>\tilde v_2$ and $v_x(0)\geq 0$ blows up at some finite point $x=x_1>0$.
The coordinate of the singularity $x=x_1$ results by setting  $v=\infty$ in the upper limit of the integral:
\begin{gather}
x_1=\int_{v_0}^\infty
\frac{\sqrt{3}d\xi}
{\sqrt{2\alpha_2(\xi^3-v_0^3)-
3(\alpha_1+1)(\xi^2-v^2_0)-6H_0(\xi-v_0)+3(v'_0)^2}}.\label{ImplSol}
\end{gather}
The integral (\ref{ImplSol}) can be bounded from above by  dropping the nonnegative term $3(v'_0)^2$ in the denominator.
Making the substitution $\xi-v_0=\eta^2$, we  obtain the upper bound:
\begin{gather}
0<x_1\leq\int_0^\infty\frac{2\sqrt{3} d\eta}
{\sqrt{ 2\alpha_2\eta^4+(6\alpha_2v_0-3\alpha_1-3)\eta^2+
6[ \alpha_2v_0^2-(\alpha_1+1)v_0-H_0]}}.\label{EstSing}
\end{gather}
Returning to the inequality (\ref{DiffIn}) we consider the equation (\ref{DiffEqV}) with  the initial data
\begin{gather*}
v(0)=\|{\bf u}_0\|^2,\quad v_x(0)=\left.\frac{d\|{\bf u}(x)\|^2}{dx}\right|_{x=0}\equiv 2({\bf u}_0,{\bf u}'_0).
\end{gather*}
The right-hand side of Eq.~(\ref{DiffEqV}) is a monotonically nondecreasing function of $v$ in the region
\[
v>v^\star\equiv
\frac{\alpha_1+1}{2\alpha_2}.
\]
Since  $\tilde v_2>v^\star$, this implies that  the right-hand side  is nondecreasing  for $v>\tilde v_2$.
The Comparison Theorem (\cite{WT}, chapter~III,  \S11, the supplement) guarantees then that $\|{\bf u}(x)\|\geq v(x)$ on any interval $I\subset \mathbb{R}^+$ where the solutions ${\bf u}(x)$ and $v(x)$ are defined.
This means that
the solution ${\bf u}(x)$ with the initial conditions
\begin{gather*}
\|{\bf u}_0\|>\tilde v_2,\quad ({\bf u}_0,{\bf u}'_0)\geq0
\end{gather*}
 blows up at some $\tilde x$, where $0<\tilde x<x_1$.
%This establishes the point (a) of the Proposition.

It is straightforward to check that if $v_0$ satisfies
\begin{gather*}
v_0>\frac1{2\alpha_2}\left(\alpha_1+1+
2\sqrt{(\alpha_1+1)^2+4\alpha_2H_0}\right),
\end{gather*}
the quartic under the radical in \eqref{EstSing} admits a simple lower bound:
\begin{gather}
2\alpha_2\eta^4+(6\alpha_2v_0-3\alpha_1-3)\eta^2+
6 \left[\alpha_2v_0^2-(\alpha_1+1)v_0-H_0\right]
\geq 2\alpha_2\left(\eta^2+\gamma^2\right)^2.
\label{inq}
\end{gather}
Here
\begin{gather*}
\gamma=\sqrt{\frac3{4\alpha_2}(2\alpha_2v_0-\alpha_1-1)}.
\end{gather*}
The lower bound for the denominator in  \eqref{EstSing}
translates into an upper bound
for the integral:
\begin{gather*}
0<x_1<\sqrt{\frac 6{\alpha_2}}\int_0^\infty\frac{d\eta}{\eta^2+\gamma^2}=
\frac{\sqrt{2}\pi}{\sqrt{2\alpha_2 v_0-\alpha_1-1}}.
\end{gather*}
This establishes the statement of Proposition 1.

In particular case, if ${\bf h}(x)\equiv 0$, inequality (\ref{DiffIn}) can be replaced by
\begin{gather*}
(\|{\bf u}\|^2)_{xx}\geq-\alpha_1\|{\bf u}\|^2+\alpha_2\|{\bf u}\|^4.
\end{gather*}
Consider instead of (\ref{DiffEqV}) the auxiliary equation
\begin{gather*}
v_{xx}=-\alpha_1v+\alpha_2v^2.
\end{gather*}
Then the formulas (\ref{EstCol2}) and (\ref{EstCol3}) follow from the Comparison Theorem in the same way as it was for (\ref{Stat03}) and (\ref{EstCol1}).
$\blacksquare$\medskip

%{\bf Remark.} It follows from the estimations above that if ${\bf %h}(x)\equiv 0$ the statement (\ref{Stat01}) holds for
%\begin{gather*}\|{\bf u}_0\|^2>M,\quad %M=\max\left(0,\frac{\alpha_1}{\alpha_2}\right),\quad  ({\bf u}_0,{\bf %u}'_0)\geq0.
%\end{gather*}

\end{document}